\documentclass[aip, jcp, superscriptaddress, showpacs, reprint, notitlepage, floatfix]{revtex4-1}
\usepackage[caption=false]{subfig}
\usepackage{graphicx,color}
\usepackage{natbib,amssymb,amsmath,amsthm,amsfonts,amsbsy,mathrsfs}
\usepackage{epsfig}
\usepackage[colorlinks   = true, urlcolor     = cyan, linkcolor    = blue, citecolor   = magenta]{hyperref}
\usepackage{algorithm}
\pdfoutput=1 
\newcommand{\Or}{\mathcal{O}}
\newcommand{\jump}[1]{\big[\hspace{-0.7mm} \big[ #1 \big]
  \hspace{-0.7mm} \big]}
\newcommand{\mean}[1] {\big\{ \hspace{-0.7mm} \big\{ #1 \big\}
  \hspace{-0.7mm} \big\}}
\newcommand{\average}[1]{\left\langle#1\right\rangle}

\newcommand{\mc}[1]{\mathcal{#1}}
\newcommand{\DG}{\mathrm{DG}}
\newcommand{\eff}{\mathrm{eff}}

\newcommand{\HDG}{H^{\mathrm{DG}}}

\newcommand{\beq}{\begin{equation}}
\newcommand{\eeq}{\end{equation}}
\newcommand{\beqs}{\begin{eqnarray}}
\newcommand{\eeqs}{\end{eqnarray}}
\newcommand{\beql}{\begin{equation} \label}

\newcommand{\half}{\frac{1}{2}}

\newcommand{\abs}[1]{\lvert#1\rvert}

\let\oldFootnote\footnote
\newcommand\nextToken\relax

\renewcommand\footnote[1]{%
    \oldFootnote{#1}\futurelet\nextToken\isFootnote}

\newcommand\isFootnote{%
    \ifx\footnote\nextToken\textsuperscript{,}\fi}

\begin{document}
\title{Chebyshev polynomial filtered subspace iteration in the Discontinuous Galerkin method for large-scale electronic structure calculations}
\author{Amartya S. Banerjee}
\thanks{asb@lbl.gov}
\affiliation{Computational Research Division, Lawrence Berkeley
National Laboratory, Berkeley, CA 94720, USA}

\author{Lin Lin}
\thanks{linlin@math.berkeley.edu}
\affiliation{Department of Mathematics, University of California,
Berkeley, CA 94720, USA} \affiliation{Computational Research
Division, Lawrence Berkeley National Laboratory, Berkeley, CA 94720,
USA}

\author{Wei Hu}
\thanks{whu@lbl.gov}
\affiliation{Computational Research Division, Lawrence Berkeley
National Laboratory, Berkeley, CA 94720, USA}

\author{Chao Yang}
\thanks{cyang@lbl.gov}
\affiliation{Computational Research Division, Lawrence Berkeley
National Laboratory, Berkeley, CA 94720, USA}

\author{John E. Pask}
\thanks{pask1@llnl.gov}
\affiliation{Physics Division, Lawrence Livermore National Laboratory, Livermore, CA 94550, U.S.A}

%% Tikz commands for flowcharts
%\tikzstyle{startstop} = [rectangle, rounded corners, line width = 0.6pt, minimum width=3cm, minimum height=1cm,text centered, draw=black, fill=red!30]
%\tikzstyle{long_startstop} = [rectangle, rounded corners, line width = 0.6pt, minimum width=4cm, minimum height=1cm, text width=4cm, text centered, draw=black, fill=red!30]
%\tikzstyle{io} = [trapezium, trapezium left angle=70, trapezium right angle=110, line width = 0.6pt, minimum width=3cm, minimum height=1cm, text centered, draw=black, fill=blue!30]
%\tikzstyle{process} = [rectangle, line width = 0.6pt, minimum width=3cm,  text width=3cm, minimum height=1cm, text centered, draw=black, fill=orange!30]
%\tikzstyle{short_process} = [rectangle, line width = 0.6pt, minimum width=2.5cm,  text width=2.5cm, minimum height=1cm, text centered, draw=black, fill=orange!30]
%\tikzstyle{long_process} = [rectangle, minimum width=4cm, line width = 0.6pt, minimum height=1cm, text centered, text width=4cm, draw=black, fill=orange!30]
%\tikzstyle{very_long_process} = [rectangle, minimum width=5cm, line width = 0.6pt, minimum height=1cm, text centered, text width=5cm, draw=black, fill=orange!30]
%\tikzstyle{decision} = [diamond, line width = 0.6pt, minimum width=3cm, minimum height=1cm, text centered, aspect=3, draw=black, fill=green!30]
%\tikzstyle{arrow} = [thick,->,>=stealth]

\date{September 15, 2016}


\begin{abstract}
{
The Discontinuous Galerkin (DG) electronic structure method employs an adaptive local basis (ALB) set to solve the Kohn-Sham equations of density functional theory (DFT) in a discontinuous Galerkin framework. 
%The methodology is implemented in the Discontinuous Galerkin Density Functional Theory (DGDFT) code for large-scale parallel electronic structure calculations. 
The adaptive local basis is generated on-the-fly to capture the local material physics, and can systematically attain chemical accuracy with only a few tens of degrees of freedom per atom. %Hence, DGDFT combines the key advantage of planewave basis sets in terms of systematic improvability with that of localized basis sets in reducing basis size.
A central issue for large-scale calculations, however, is the computation of the electron density (and subsequently, ground state properties) from the discretized Hamiltonian in an efficient and scalable manner. We show in this work how Chebyshev polynomial filtered subspace iteration (CheFSI) can be used to address this issue and push the envelope in large-scale materials simulations in a discontinuous Galerkin framework. 
%In particular, this strategy makes it possible to attack complex materials problems involving many thousands of atoms routinely. 
We describe how the subspace filtering steps can be performed in an efficient and scalable manner using a two-dimensional parallelization scheme, thanks to the orthogonality of the DG basis set and block-sparse structure of the DG Hamiltonian matrix. The on-the-fly nature of the ALBs requires additional care in carrying out the subspace iterations. We demonstrate the parallel scalability of the DG-CheFSI approach in calculations of large-scale two-dimensional graphene sheets and bulk three-dimensional lithium-ion electrolyte systems. Employing $55,296$ computational cores, the time per self-consistent field iteration for a sample of the bulk 3D electrolyte containing $8,586$ atoms is $90$ seconds, and the time for a graphene sheet containing $11,520$ atoms is $75$ seconds.
}
\end{abstract}

\maketitle
\section{Introduction}
Kohn-Sham Density functional theory (KS-DFT) \cite{HK_DFT, KohnSham_DFT} is the most widely used methodology for electronic structure calculations of condensed matter and nano-material systems. KS-DFT requires the solution of a nonlinear eigenvalue problem, and this is usually achieved by means of self-consistent field (SCF) iterations in conjunction with convergence acceleration schemes \cite{Martin_ES, Kohanoff}. The most computationally intensive part of conventional KS-DFT calculations is the solution of the linear eigenvalue problem associated with diagonalization of the Kohn-Sham Hamiltonian on every SCF step. The results of this eigenvalue problem are used to update the electron density $\rho$, from which the various terms of the Kohn-Sham Hamiltonian are computed. As the SCF iterations progress, the solution to the linear eigenvalue problem on successive SCF steps forms increasingly better approximations to the actual Kohn-Sham eigenstates.

The computational complexity {(or algorithmic complexity)} of the solution of the linear eigenvalue problem, with respect to the number of electronic states in the system, is dependent on the algorithm used for solution of the eigenvalue problem -- in particular, it depends on whether direct or iterative methods of solution are used. The prefactor in such algorithmic complexity estimates strongly influences the simulation wall times in practical computations. The ability to tackle large-scale complex materials science problems, therefore, is closely related to how small the prefactor can be made in real computations, regardless of which diagonalization algorithm is used.

The prefactor is not only influenced by the choice of algorithm, but also by the discretization scheme. Specifically, it depends on the number of basis functions per atom required to obtain accurate and reliable results. The widely used planewave method \cite{Teter_Payne_Allan_2, Martin_ES, Kresse_abinitio_iterative} allows high fidelity calculations to be carried out since systematic convergence with respect to the number of basis functions per atom can be obtained. However, this method requires a large number of basis functions per atom -- often thousands or more planewaves per atom need to be employed. Similar observations hold true for methods based on finite elements \cite{Pask_FEM_review, tsuchida1995electronic, chen2014adaptive, Gavini_Kohn_Sham}, finite differences \citep{Chelikowsky_Saad_1, Chelikowsky_Saad_3, ghosh2016sparc, Octopus_1}, and other planewave-like spectral basis functions \citep{banerjee2015spectral, My_PhD_Thesis}. On the other hand, methods based on atom centered basis functions  \citep{LCAO_3, LCAO_famous, SIESTA_1, blum2009ab} typically require fewer basis functions per atom (often, as few as 10 -- 80). However, it can be nontrivial to improve the quality of solutions obtained via such methods, as a result of which the success of a practical calculation can depend on the experience of the practitioner.

In a series of recent contributions \citep{lin2012adaptive, zhang2015adaptive, hu2015dgdft, hu2015edge}, a new methodology for discretizing the Kohn-Sham equations using so called adaptive local basis functions (ALBs) has been presented.  This approach involves partitioning the global simulation domain into a set of subdomains (called elements) and solving the Kohn-Sham equations locally in and around each element. The results of these local calculations are used to generate the ALBs (in each element) and the Kohn-Sham equations in the global simulation domain are then discretized using them. Since the ALBs form a discontinuous basis set globally (the discontinuity occurs at the element boundaries), the interior penalty Discontinuous Galerkin (DG) approach~\cite{Arnold1982} is used for constructing the Hamiltonian matrix. The DG formulation ensures that the global continuity of the relevant Kohn-Sham eigenstates and related quantities such as electron density is approximately obtained.
 
The solution obtained by the above procedure converges systematically to the infinite basis set limit as the number of ALBs is increased. The error in this scheme can be gauged by means of \textit{a posteriori} error estimators \citep{dgdft_posteriori,lin2015posteriori,lin2016posteriori}. Owing to the fact that the ALBs incorporate local materials physics into the basis, an efficient discretization of the Kohn-Sham equations can be obtained in which chemical accuracy in total energies and forces can be attained with a few tens of basis functions per atom \citep{lin2012adaptive, zhang2015adaptive}. The DG approach for  solving the Kohn-Sham equations with ALBs has been incorporated into a massively parallel software package called DGDFT \citep{hu2015dgdft, hu2015edge}.

Although the DG framework for the Kohn-Sham equations (as implemented in the DGDFT code) has been successfully used to study materials problems involving many thousands of atoms \citep{hu2015edge}, a persistent issue has been to obtain the electron density from the discretized Kohn-Sham Hamiltonian in an efficient manner for systems containing a thousand atoms or more.  Due to the relatively small size of the discretized Hamiltonian matrices involved,  direct diagonalization methods (via ScaLAPACK \citep{blackford1997scalapack, choi1995scalapack}, for example) are feasible for systems of smaller size. However, the computational cost of the these methods scales in a cubic manner, i.e., as $\Or(N_b^3)$, with $N_b$ denoting the total number of basis functions used in the simulation. Thus, the computational cost increases steeply with respect to the size of the system. The cubic scaling problem is compounded by the fact that direct diagonalization solvers (for dense matrices) typically do not scale well beyond a few thousand processors on distributed memory machines. In recent work with the DGDFT code in massively parallel computing environments \citep{hu2015dgdft, hu2015edge}, we have found that for systems containing more than a few thousand atoms, the step of obtaining the electron density from the Hamiltonian can consume $95\%$ or more of the total computational time. 

Direct diagonalization methods do not take advantage of the fact that the DG Hamiltonian matrix, denoted henceforth by $H^{\textrm{DG}}$, is a block-sparse matrix. The sparsity of $H^{\textrm{DG}}$  allows several alternatives for mitigating the issues associated with direct diagonalization methods. One alternative is to employ ``linear scaling'' methods (see, e.g.,~review articles \citep{goedecker1999linear, bowler2012methods}), based on direct calculation of truncated density matrices. While linear scaling methods have been very successful for tackling large insulating systems with sizable band gaps~\citep{hine2009linear, vandevondele2012linear, fattebert2006linear}, they are less well suited for metallic systems or semiconducting systems with small band gaps. The sparse nature of the $H^{\textrm{DG}}$ matrix also allows for the Pole Expansion and Selected Inversion (PEXSI) technique \citep{CMS2009, JPCM_25_295501_2013_PEXSI} to be employed for directly computing the electron density and other ground state properties. The computational cost of the PEXSI technique is at most $\Or(N_b^3/N_s)\sim \Or(N_s^2)$, with $N_s$ denoting the number of Kohn-Sham states, even for metallic systems. The PEXSI technique has excellent parallel scalability \citep{lin2014siesta, hu2015edge, hu2015dgdft} and has been shown to work well in conjunction with DGDFT while studying two-dimensional materials. However, it becomes more expensive (both in terms of memory and run time) for three-dimensional bulk materials, and has limited ability to make use of good starting guesses (from previous geometry optimization or molecular dynamics steps) to accelerate computations.

With the above considerations in mind, an alternate strategy for reducing the simulation wall time in practical computations is to revert to the usage of an algorithm that scales in a cubic manner with respect to the system size, but to reduce the pre-constant of the algorithm.  This includes the use of iterative diagonalization methods such as the Davidson method \citep{dav75,dav89}, 
conjugate gradient-type methods \citep{Teter_Payne_Allan_2, ABINIT_LOBPCG, vecharynski2015projected}, and residual minimization methods~\citep{Kresse_abinitio_iterative}. However, the effectiveness of these schemes relies on the availability of a good preconditioner, which is currently not available for $H^{\textrm{DG}}$.

In this work, we utilize the technique of Chebyshev polynomial filtered subspace iteration (CheFSI) to address the diagonalization problem in the DG framework and implement it within the DGDFT code. While the CheFSI technique has been utilized with great success by various practitioners working with finite differences, finite elements, and spectral basis sets \citep{Serial_Chebyshev, Parallel_Chebyshev, zhou_2014_chebyshev, michaud2016rescu, Gavini_higher_order, banerjee2015spectral, CyclicDFT_JMPS, levTor15}, its application to basis sets resulting in reduced-size Hamiltonian matrices (such as atomic orbital type or adaptive basis sets), with on-the-fly adaptation in particular, has not been considered before to our knowledge.

The DG framework has a number of features that make the use of CheFSI attractive. First, since the ALBs are orthonormal, one does not need to consider the overlap matrix (for usual pseudopotential calculations), thus ensuring that the CheFSI method in its original form can be readily employed. Secondly, {compared to Hamiltonian and overlap matrices resulting from other high-quality orbital-based basis sets, such as augmented Gaussians \citep{rappoport2010property} or partition-of-unity finite-elements \citep{cai2013hybrid}, the $H^{\textrm{DG}}$ matrix has relatively small spectral radius -- of the order of a few thousand (atomic units) for the systems considered here. As a result of this, Chebyshev polynomial filters of relatively low order suffice.}  In contrast to direct diagonalization methods which scale as $\Or(N_b^3)$, the computational complexity of CheFSI \citep{Serial_Chebyshev} is reduced to $\Or(N_b N_s^2 + N_s^3)$. Since $N_b/N_s$ is typically $\sim 2 - 20$ for ALBs, this reduction of prefactor can be sizable for systems with thousands of atoms, leading to substantially shorter simulation times. Finally, due to the sparse nature of the $H^{\textrm{DG}}$ matrix and its nearest neighbor block structure, the computation of the product of this matrix with a block of dense vectors can be carried out with relatively low communication volume between processors. This observation leads to an efficient and scalable Hamiltonian matrix times vector product implementation that is crucial for the success of the CheFSI method within the DG framework.

Overall, we seek an approach for conventional KS-DFT calculations of large systems with substantially reduced prefactor, while retaining the accuracy, systematic improvability, and general applicability of established planewave and other spectral approaches. This is made possible by the use of ALBs which ensure that the number of degrees of freedom per atom is kept low, combined with the use of the CheFSI method which is known to have a low prefactor compared to conventional algorithms -- as long as an efficient implementation of the Hamiltonian matrix times vector product can be set up. As we show subsequently, through this combination of strategies, we are able to tackle systems containing thousands of atoms routinely, with wall times on the order of a few tens of seconds per SCF step {on large-scale parallel computing clusters}.

The rest of the paper is structured as follows. In Section \ref{sec:methodology}, we outline the background on the DG formulation of KS-DFT and the CheFSI method, before delving into the implementation of the CheFSI method within the DG framework. In Section \ref{sec:results}, we present results and comparisons with competing methods. We conclude and comment on future research directions in Section \ref{sec:conclusion}. 
%\vspace{-0.5cm}
\section{Methodology}
\label{sec:methodology}
%\vspace{-0.25cm}
\subsection{Discontinuous Galerkin formulation of KS-DFT}
In this section, we discuss aspects of the DG framework for the Kohn-Sham equations -- as implemented within the DGDFT code -- relevant to the implementation of the CheFSI method. More details on the theoretical underpinnings and practical implementation strategies of the DG framework can be found in \citep{hu2015dgdft, lin2012adaptive, zhang2015adaptive}. 

\begin{figure}
\subfloat[Schematic $4 \times 4$ partition of a model 2D computational domain into 16 elements.\label{subfig:partition}]{
  \includegraphics[width=0.4\linewidth]{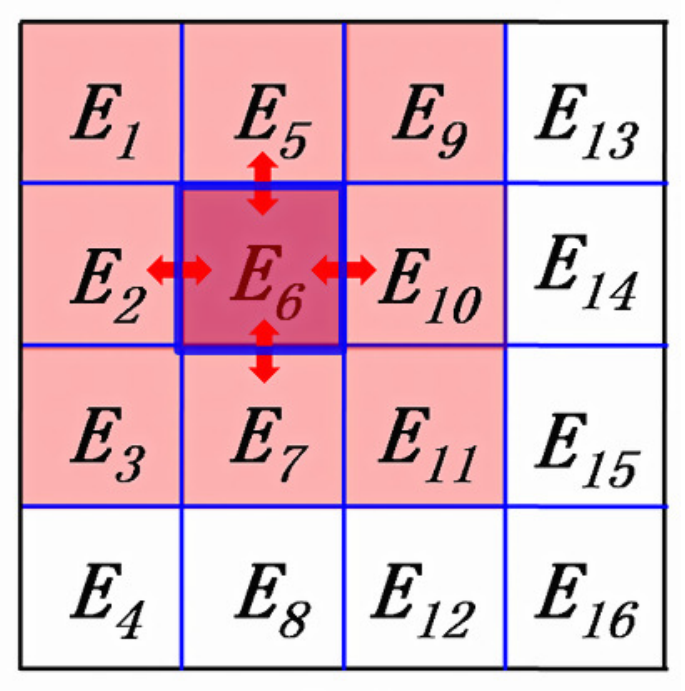}
}\\
\subfloat[The DG Hamiltonian matrix $H^{\text{DG}}$ with its block sparsity pattern resulting from such a partition. White represents zero blocks. 
\label{subfig:HDG}]{
  \includegraphics[width=0.8\linewidth]{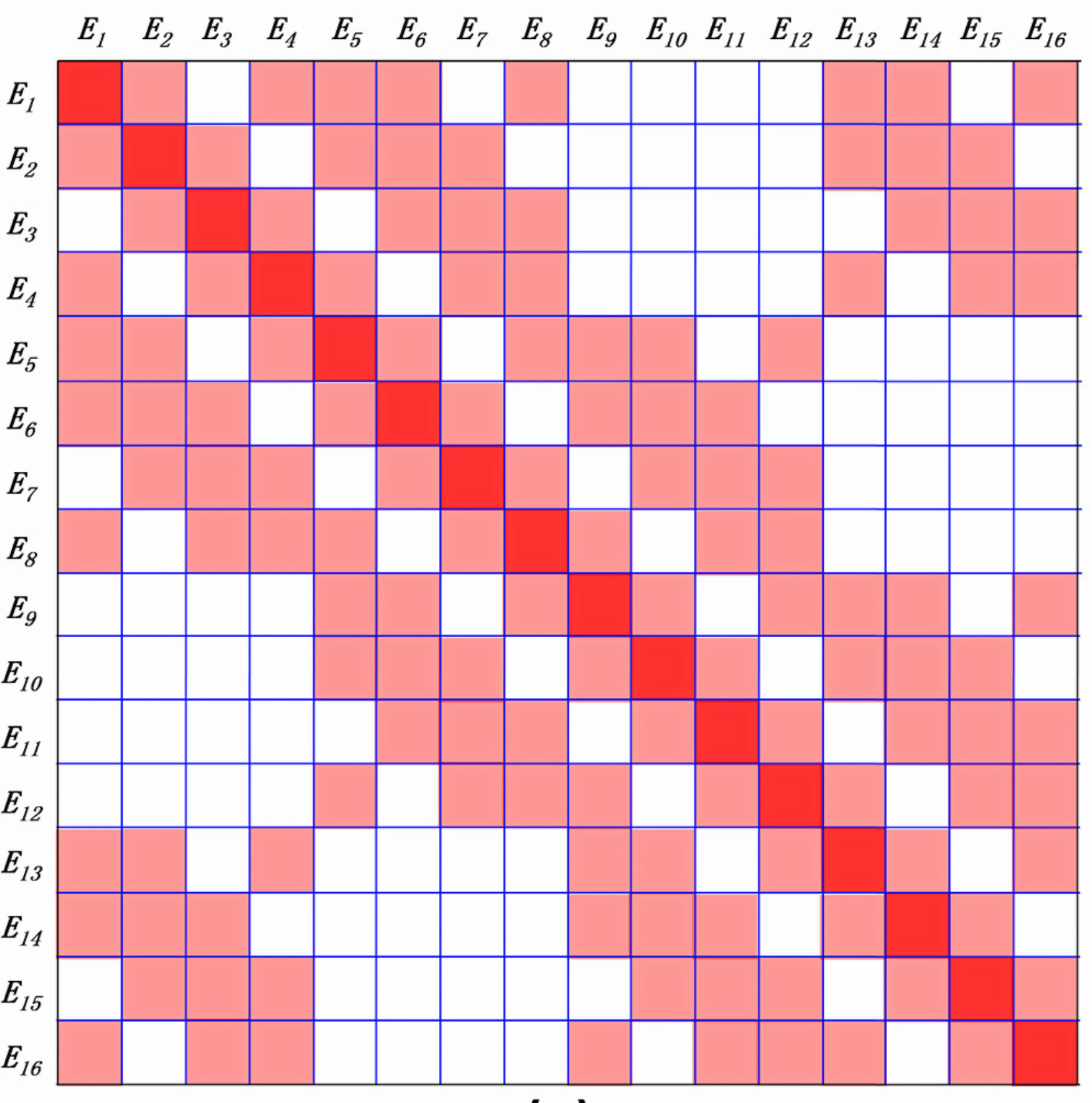}
}
\caption{Partitioning of a domain into DG elements and the resulting discretized Hamiltonian $H^{\text{DG}}$.}
\label{fig:Partition_and_HDG}
\end{figure}

In the present work, we consider $\Gamma$-point calculations of periodic systems, as typical in \textit{ab initio} molecular dynamics, and large-scale calculations generally. The Kohn-Sham orbitals can be taken to be real valued in this case. In the DG framework, the global simulation domain $\Omega$ is partitioned into a number of subdomains (or \emph{elements}) such that the union of these subdomains tile the whole domain and adjacent subdomains are non-overlapping (except possibly at corners, edges, or at a surface). Due to the periodic boundary conditions on $\Omega$, each surface of each element is shared between two neighboring elements. We denote the collection of the sub-domains as $\mc{T} = \{E_{1},\ldots,E_{M}\}$ and the collection of all the surfaces as $\mc{S}$.  Each element $E_{K}$ is embedded into a slightly larger \emph{extended element} $Q_{K}$ that includes a buffer region surrounding $E_{K}$. Figure \ref{subfig:partition} shows a model 2D system partitioned using 16 equal elements $\{E_{1},\ldots,E_{16}\}$. 

Due to the decomposition of $\Omega$ into elements, global $L^{2}$ inner products between various quantities (denoted here as $\average{\cdot, \cdot}_{\mc{T}}$), can be taken as the sum of local $L^{2}$ inner products over individual elements. We introduce the notation $\average{\cdot, \cdot}_{\mc{S}}$ to define the sum of local $L^{2}$ surface inner products  on all surfaces of all elements. We will also employ the notation $\mean{\cdot}$ to denote the average of a quantity across a surface while $\jump{\cdot}$  denotes the jump across a surface for discontinuous quantities.

In DGDFT, each SCF iteration includes a preliminary step of generating the ALBs on the fly. This is accomplished by iteratively solving a local Kohn-Sham problem on each of the extended elements -- the effective potential used for this calculation is simply the restriction of the effective potential on the global simulation domain to the extended element. The resulting (approximate) Kohn-Sham states over $Q_{K}$ are then restricted to $E_{K}$ and orthonormalized to produce the ALBs over $E_{K}$. 

At the end of the ALB generation process, each element $E_{K}$ has a collection of $J_K$ ALBs, denoted by $\{\varphi_{K,j}\}_{j=1}^{J_K}$. Each ALB is compactly supported on one element. The complete collection of ALBs
\begin{align}
\label{eq:ALB_set}
\mathcal{A} = \big\{\varphi_{K,j} \big \}_{K=1,j=1}^{K=M, j=J_K}\,,
\end{align}
forms an $L^{2}$ orthonormal set over $\Omega$, i.e., 
\begin{align}
 \average{\varphi_{K, j}, \varphi_{K', j'}}_{\mc{T}} = \delta_{K,K'}\delta_{j,j'}\,,
\label{eq:ALB_orthonormal}
\end{align}
for $K,K'=1,\ldots,M;\;j=1,\ldots,J_{K}\; \textrm{and}\; j'=1,\ldots,J_{K'}$.
The global Kohn-Sham states over $\Omega$ can be expanded using the ALBs as:
\begin{align}
  \psi_i(x) = \sum_{K=1}^{M} \sum_{j=1}^{J_K} c_{i; K, j}
  \varphi_{K, j}(x)\,.
 \label{eq:psiexpand}
\end{align}
Due to the fact that the ALBs are discontinuous over the global domain, whereas the Kohn-Sham states (and related physical quantities such as the electron density) are continuous, the DG framework penalizes discontinuities in these quantities across element surfaces. Accordingly, the electronic free energy of a system with $N_s$ occupied electronic states is written as: 
\begin{align}
\nonumber
E^{\DG}_{\textrm{free}}(\{\psi_i,f_i\}) &= \half \sum_{i=1}^{N_s} 2f_i\average{\nabla
    \psi_i , \nabla \psi_i}_{\mc{T}}
    + \average{ V_{\eff}, \rho }_{\mc{T}}  \\\nonumber
    &+ \sum_{i=1}^{N_s} 2f_i \sum_{I=1}^{N_A} \sum_{\ell=1}^{L_{I}} \gamma_{I,\ell} 
    \abs{\average{b_{I,\ell}(\cdot-R_{I}), \psi_i}_{\mc{T}}}^2 \\\nonumber
    &- T_{\textrm{el}}\,S_{\textrm{el}}(\{f_i\})
    \\\nonumber
    &- \sum_{i=1}^{N_s}
    2 f_i \average{\mean{\nabla\psi_i}, \jump{\psi_i}}_{\mc{S}}\\
    &+ \alpha \sum_{i=1}^{N_s} 2 f_i \average{\jump{\psi_i},
    \jump{\psi_i}}_{\mc{S}}\,,  
\label{eq:KS_DG_Energy}
\end{align}
where, $0 \leq f_i \leq 1$ are the electronic occupation numbers (specified via Fermi-Dirac smearing), $V_{\eff}$ denotes the effective potential (consisting of local pseudopotential, Hartree, and exchange correlation contributions), the scalars $\gamma_{I,\ell}$ and projector functions $b_{I,\ell}$ correspond to the nonlocal pseudopotential expressed in the Kleinman-Bylander form \citep{PRL_48_1425_1982}, and $T_{\textrm{el}}$ and $S_{\textrm{el}}$ correspond to the electronic temperature and electronic entropy, respectively. The quantity $\alpha$ is an adjustable penalty parameter that ensures that Eq.~\eqref{eq:KS_DG_Energy} has a well defined
ground state free  energy.

Using the ALBs to discretize the above expression for the free energy and subsequently minimizing the discretized energy with respect to the expansion coefficients $c_{i; K, j}$ (as well as the occupation numbers $f_i$), while maintaining the orthonormality constraint on the orbitals, leads us to the discretized version of the Euler--Lagrange equations. This takes the form of the following eigenvalue problem:
\begin{align}
\sum_{K',j'} H^{\DG}_{K, j;  K', j'} c_{i; K',j'} = \lambda_{i} c_{i;K,j}\,,
  \label{eq:eigvalue_prob}
\end{align}
with the discretized Hamiltonian operator expressible as:
\begin{align}
\label{eq:HamDG}
\nonumber
&H^{\DG}_{K, j;  K', j'} \\\nonumber
=&\;\Bigl(\frac{1}{2} \average{\nabla
    \varphi_{K, j}, \nabla \varphi_{K, j'}}_{\mc{T}}
    + \average{\varphi_{K, j}, V_{\eff}\varphi_{K, j'}}_{\mc{T}}
    \Bigr) \delta_{K,K'}\\\nonumber
    &+ \Bigl(\sum_{I,\ell} \gamma_{I,\ell} \average{\varphi_{K, j},
    b_{I,\ell}}_{\mc{T}} \average{b_{I,\ell}, \varphi_{K', j'}}_{\mc{T}}
    \Bigr) \\\nonumber
    &+ \Bigl(- \frac{1}{2} \average{\jump{\varphi_{K, j}},
    \mean{\nabla\varphi_{K', j'}}}_{\mc{S}} \\\nonumber
    &\phantom{+ \Bigl(} - \frac{1}{2} \average{\mean{\nabla\varphi_{K, j}}, \jump{\varphi_{K',j'}}}_{\mc{S}}\\
    &\phantom{+ \Bigl(} + \alpha \average{\jump{\varphi_{K, j}}, \jump{\varphi_{K',
    j'}}}_{\mc{S}}\Bigr)\,.
\end{align}
The matrix $H^{\DG}$ can be naturally partitioned into blocks based on the element indices. We will denote the $(K,K')$-th matrix sub-block (of size $J_{K}\times J_{K'}$) as:
\begin{align}
\label{eq:HDG_sub_blocks}
H^{\DG}_{K;K'} = H^{\DG}_{K,j = 1,\ldots, J_{K};K', j' = 1,\ldots, J_{K'}}\,.
\end{align}
Since the ALBs are compactly supported on their respective elements, the block $H^{\DG}_{K;K'}$ is non zero only when $K$ and $K'$ refer to neighboring elements. This situation is illustrated in Figure \ref{fig:Partition_and_HDG}.

Further details on interpretation and computation of the various terms described above as well as the significance of the parameter $\alpha$  in practical calculations can be found in \citep{hu2015dgdft, lin2012adaptive}. Note that the appearance of average and jump operators in Eq.~\eqref{eq:KS_DG_Energy} and Eq.~\eqref{eq:HamDG} are a distinguishing feature of the interior penalty DG formulation of the Kohn-Sham equations.

During the SCF iterations, the matrix $H^{\DG}$ is constructed using the most recent effective total potential $V_{\textrm{eff}}$. Following this, the electron density needs to be computed from $H^{\DG}$. So far, this step has been achieved in the DGDFT code in two distinct ways. In the first approach, the use of a parallel eigensolver (the ScaLAPACK routine PDSYEVD) allows one to directly compute the eigenvalues and eigenvectors of $H^{\DG}$. From these, the orbital occupations $\{f_i\}_{i=1}^{N_s}$ can be computed via Fermi-Dirac smearing while the density matrix (also called the \textit{Fermi matrix} at finite electronic temperature) and the electron density can be computed from the eigenvectors as:
\begin{align}
\label{eq:densitymatrix}
 P_{K,j;K',j'} &= \sum_{i=1}^{N_s} f_i c_{i;K,j} c_{i;K',j'}\,,\\
\label{eq:electron_density_from_DM}
\rho(x) &= 2\sum_{K=1}^{M} \sum_{j=1}^{J_{K}} \sum_{j'=1}^{J_{K}}
  \varphi_{K,j}(x) \varphi_{K,j'}(x) P_{K,j;K,j'}\,.
\end{align}
Eq.~\eqref{eq:electron_density_from_DM} shows that only the diagonal blocks of the density matrix need to be computed to evaluate the electron density. Note that the calculation of these blocks can be done individually on each element.

The second approach involves the use of the PEXSI technique to directly compute the density matrix elements, without going through the intermediate eigenvalues and eigenvectors. The electron density can be evaluated subsequently from Eq.~\eqref{eq:electron_density_from_DM} using the diagonal blocks of the density matrix, while the computation of forces requires computation of the non-diagonal blocks corresponding to the sparsity pattern of $H^{\DG}$\citep{lin2014siesta,zhang2015adaptive}.

Considering the limitations of the each of the above approaches (as described earlier), we now explore the option of using Chebyshev polynomial filtered subspace iteration to compute the eigenvalues and eigenvectors of $H^{\DG}$.
%\vspace{-0.5cm}
\subsection{Chebyshev polynomial filtered subspace iteration within DGDFT}
\label{subsec:cheby_within_DG}
%\vspace{-0.5cm}
Subspace iteration is a generalization of the classical power method for computing the dominant eigenpair of a matrix \citep{Saad_large_eigenvalue_book, Online_Template_Eigenvalue_Problem_Book}. The standard subspace iteration can be used to obtain an approximation to the invariant subspace associated with the largest few eigenvalues. In Kohn-Sham DFT, the invariant subspace of interest is the one associate with the occupied states (and possibly a few unoccupied states above the Fermi level) \citep{stephan1998improved, baroni1992towards} which do not correspond to the dominant eigenvalues of the Kohn-Sham Hamiltonian.

A Chebyshev polynomial $p_m(\lambda)$ can be constructed to map eigenvalues at the low end of the spectrum ({corresponding to the occupied states}) of $H^{\DG}$ to the dominant eigenvalues of $p_m(\HDG)$. The exponential growth property of {the Chebyshev} polynomials outside the region $[-1,1]$ can be used to ensure that the wanted part of the spectrum (i.e., the occupied states in ground state electronic structure calculations) can be magnified while the unwanted part ({corresponding to the unoccupied states}) is damped in comparison \citep{bauer1957verfahren,rutishauser1969computational, rutishauser1970simultaneous}. Applying subspace iteration to $p_m(\HDG)$ yields the desired invariant subspace. Within each iteration, the multiplication of $p_m(\HDG)$ with a block of vectors can be carried out by using the three-term recurrence satisfied by Chebyshev polynomials. The application of the Chebyshev polynomial filtered subspace iteration (CheFSI) technique for computing the occupied eigenspace of the Kohn-Sham operator was introduced in \cite{Serial_Chebyshev, Parallel_Chebyshev}. Within the SCF iteration framework, this methodology can be thought of as a form of nonlinear subspace iteration in the sense that the approach de-emphasizes the accurate solution of the intermediate linearized Kohn-Sham eigenvalue problems on every SCF step.  With the progress of the SCF iterations, the approximate Hamiltonian approaches the self-consistent one and the span of the (approximately) computed eigenvectors approaches the converged occupied subspace simultaneously. This particular feature of CheFSI has some bearing on the way it is implemented within the DG framework, as explained later.

The main desirable features of CheFSI which make it suitable for application to large-scale electronic structure problems are the following : 1) It is a block method in which $\HDG$ can be multiplied with a block of vectors simultaneously. This additional level of concurrency allows the algorithm to achieve better parallel scalability compared to standard Krylov subspace methods such as the Lanczos algorithm. 2) Compared to other Krylov subspace methods, it performs fewer Rayleigh-Ritz calculations in which a projected subspace eigenvalue problem is solved. The Rayleigh-Ritz procedure is often the computational bottleneck when the number of eigenvalues to be computed is relatively large. 

The key steps in a CheFSI cycle (see Algorithm \ref{algo1} for details) are an application of the Chebyshev polynomial filter  on a block of vectors, subsequent orthonormalization of the filtered block, a Rayleigh-Ritz step, and finally a so called subspace rotation step \citep{zhou_2014_chebyshev,Serial_Chebyshev}.  Together, the last three steps will be referred to as solving the \emph{subspace problem}. Note that the Rayleigh-Ritz and subspace rotation steps are useful for explicitly obtaining the (approximate) occupied eigenpairs of the Hamiltonian from the filtered subspace. We will now elaborate on various important aspects of our implementation of these steps within DGDFT.
\begin{algorithm}[H]
\caption{CheFSI cycle}
{\bf Input:} Matrix $\HDG$, starting vector block $X$, filter order $m$
\begin{enumerate}
\item Compute lower bound $b_\text{low}$ using previous Ritz values and the upper bound $b_\text{up}$ using a few steps of the Lanczos algorithm.
\item Perform Chebyshev polynomial filtering, i.e., compute $\tilde{Y} = p_m(\HDG) X$ with $[b_\text{low}, b_\text{up}]$ mapped to $[-1,1]$.
\item Orthonormalize columns of $\tilde{Y}$: Set $S = \tilde{Y}^T \tilde{Y}$, compute $U^TU = S$, and  solve $\hat{Y}U = \tilde{Y}$.
\item Rayleigh-Ritz step: Compute the projected subspace matrix $\hat{H} = \hat{Y}^{T}\HDG \hat{Y}$ and solve the eigenproblem $\hat{H} Q = Q D$.
\item Perform a subspace rotation step $X_{\textrm{new}} = \hat{Y} Q$.
\end{enumerate}
{\bf Output:} Vector block $X_{\textrm{new}}$ (approximate eigenvectors) and Ritz values $D$ (approximate eigenvalues). 
\label{algo1}
\end{algorithm}

\subsubsection{The multiplication of \texorpdfstring{\emph{$H^{\text{DG}}$}}{} with a block of vectors}
\label{subsubsec:matvec}
One of the key computational steps of the CheFSI method is to perform $Y = p_m(H^{\DG})\,X$. This step requires an efficient and scalable implementation of multiplying $\HDG$ with a block of vectors. Additionally, computation of the action of the Hamiltonian matrix on vectors is required for estimating the spectral bounds of the Hamiltonian via the Lanczos algorithm as well as during the Rayleigh-Ritz step (see Algorithm \ref{algo1}).

Given a block of vectors $X$ consisting of columns $\{x_i\}_{i=1}^{N_s}$, the multiplication of $\HDG$ with a subset of these columns $x_{i_1,\ldots,i_2}$, with $1 \leq i_1 \leq i_2 \leq N_s$, can be written as:
\begin{align}
\label{eq:matvec_1}
y_{i_1,\ldots,i_2;K, j} = \sum_{K',j'} H^{\DG}_{K, j;  K', j'} x_{i_1,\ldots,i_2;K',j'}\,,
\end{align}
where, $K,K'=1,\ldots,M$ and $j=1,\ldots,J_K, j'=1,\ldots,J_{k'}$. Using the fact that $H^{\DG}_{K;K'}$ is non-zero only for $K' \in \mathcal{N}(K)$, i.e., the neighboring elements of the $K$-th element, we may rewrite this as:
\begin{align}
\label{eq:matvec_2}
y_{i_1,\ldots,i_2;K, j} = \sum_{K'\in \mathcal{N}(K)}H^{\DG}_{K;  K'} x_{i_1,\ldots,i_2;K', j'}\,.
\end{align}
Thus, the portion of the resulting set of vectors $y_{i_1,\ldots,i_2}$ that is associated with the element $K$ can be written as:
\begin{align}
\label{eq:matvec_3}
y_{i_1,\ldots,i_2;K} = \sum_{K'\in \mathcal{N}(K)}H^{\DG}_{K;  K'} x_{i_1,\ldots,i_2;K'}\,.
\end{align}
Since the individual blocks $ H^{\DG}_{K;  K'}$ and $x_{i_1,\ldots,i_2;K'}$ are dense,  Eq.~\eqref{eq:matvec_3} can be computed as the sum of a series of matrix-matrix products (i.e., GEMM operations in Level-3 BLAS). Further, since the above operation can be carried out independently over the various columns of $X$, it is natural to take advantage of the manifestly parallel nature of the problem by distributing the columns among separate processing elements in an appropriate manner.

\begin{figure}
\subfloat{\includegraphics[width=0.75\linewidth]{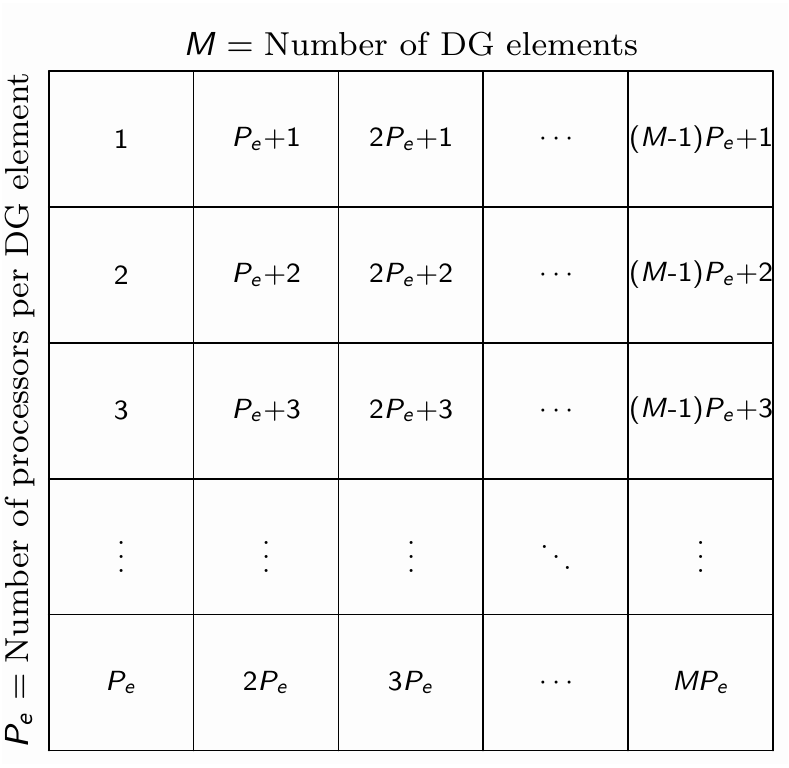}
}
\caption{Two-level parallelization scheme within DGDFT : The number of DG elements is $M$ and each element has $P_{e}$ processors dedicated to it, making the total number of processors $P_{\mathrm{tot}} = M \times P_{e}$.}
\label{fig:parallelization}
\end{figure}

The data distribution of the various quantities involved in Eq.~\eqref{eq:matvec_3} is important in deciding how the operation can be carried out in practice. As explained in \citep{hu2015dgdft}, the DGDFT code uses a two level parallelization strategy implemented via Message Passing Interface (MPI) to handle inter-process communication. At the coarse grained level, work is distributed among processors by elements, leading to \textit{inter-element} parallelization. Further, within each element, the work associated with construction of the local portions of the DG Hamiltonian, evaluation of the electron density, and the ALB generation process is  parallelized leading to \textit{intra-element} parallelization. The processors are partitioned into a two-dimensional logical process grid with a column major order (Fig \ref{fig:parallelization}).  We will refer to this layout of the MPI processes as the \emph{global process grid}.
 For the sake of discussion, we assume here that the total number of MPI processes (in the global process grid) is $P_{\mathrm{tot}} = M \times P_{e}$, so that there are $P_e$ processes assigned to each of the $M$ elements. Specifically, the processes with MPI ranks $(K-1)P_{e}+1$ to $K P_{e}$ ($K=1,\ldots, M)$ are in the $K$-th global column processor group, and they work on the element $E_{K}$ at the level of intra-element parallelization. Analogously, the $i$-th global row processor group consists of the processes with MPI ranks $i,P_{e}+i,\ldots,(M-1)P_{e}+i$ ($i=1,\ldots,P_{e}$). Any process in the global process grid can be referred to by its row and column indices $(I,K)$ in the process grid, with $I=0,\ldots, P_e-1$ and $K=0,\ldots,M-1$.

The two level parallelization scheme within DGDFT allows \eqref{eq:matvec_3} to be evaluated efficiently and in a scalable manner. A given block of vectors $X$ is distributed at the inter-element parallelization level in a manner consistent with the element-wise partition suggested by Eq.~\eqref{eq:matvec_3}, and it is further distributed by its columns (i.e., Kohn-Sham states) at the intra-element parallelization level. In other words, for a block of vectors of size $N_b \times N_s$ (with $N_b = \sum_{k=1}^{M} J_K$, i.e., the total number of basis functions in use and $N_s$ denoting the total number of Kohn-Sham states), the MPI process with row and column indices $(I,K)$ holds a block of size $J_K \times \lfloor\frac{N_s}{P_e}\rfloor$ and evaluates this portion of the result (i.e, left hand side of Eq.~\eqref{eq:matvec_3}). In DGDFT, the matrix $H^{\DG}$ is stored element-wise, i.e., all $P_e$ processes in a given process grid column assigned to a particular element $K$ store all non-zero blocks of the form $H^{\DG}_{K;  K'}$. Hence, for a given process, evaluation of \eqref{eq:matvec_3} only incurs nearest neighbor communication within each process grid row so that blocks of the form $x_{i_1,\ldots,i_2;K'}$ (with $i_1, i_2$ corresponding to the start and end indices of the block of states that the process is working on, and $K'$ corresponding to its nearest neighbor elements) may be obtained.

The strategy of employing a process grid avoids costly global communication and restricts all communication to individual row and column process grids. Additionally, the block nearest-neighbor type sparsity structure of the Hamiltonian matrix results in further reduction in communication volume. As demonstrated later, these factors result in a particularly well scaling matrix-vector product routine for DGDFT. In contrast, expressing $\HDG$ and the block of vectors to be multiplied as dense matrices and the subsequent direct use of parallel dense linear algebra routines (PBLAS \citep{choi1995proposal}, for example) for carrying out the matrix-vector product operation would have incurred a higher computational cost and also significantly degraded the scalability of the computation. 
\label{subsubsec:MATVEC}
\subsubsection{Parallel solution of the subspace problem}
\label{subsubsec:ScalaPACK_Sol}
The various steps involved in solving the subspace problem all require dense linear algebra operations. For example, given the Chebyshev polynomial filtered block of vectors $\tilde{Y} = p_m(H^{\DG})\,X$, we need to orthonormalize this block of vectors so as to obtain an orthonormal basis for the (approximate) occupied subspace. We carry out this operation by computing the overlap matrix $S = \tilde{Y}^T \tilde{Y}$, computing the Cholesky factorization of $S$ as $S = U^TU$ and then using the Cholesky factor to solve the equation $\hat{Y}U = \tilde{Y}$. The resulting block of vectors $\hat{Y}$ is then orthonormal.  The cost of these operations grows cubically with respect to the number of atoms involved in the simulation. Once the number of occupied states exceeds a few hundred, it becomes necessary to parallelize these operations so as to reduce the computational wall times.  We use the parallel dense linear algebra routines in the PBLAS \citep{choi1995proposal} and  ScaLAPACK \citep{blackford1997scalapack, choi1995scalapack} software libraries to do this.

PBLAS and ScaLAPACK routines employ a two-dimensional block-cyclic data distribution over a process grid for their operations. We will refer to this process grid as the \emph{linear algebra process grid}. Since the performance of some of the required routines (particularly, those involved with eigenvalue computation and Cholesky factorization) tend to stagnate (or sometimes, even deteriorate) quite easily if too many processes are in use, we typically use only the first row of processes of the global process grid to set up the linear algebra process grid for problems of moderate size. Thus, the number of processes in the linear algebra process grid typically equals the number of DG elements in use. As the system size grows bigger, we include additional rows of processors in the global processor grid in the linear algebra process grid to reduce the cost of dense linear algebra operations. 

Before the sequence of dense linear algebra operations can be initiated, the vector block that contains the product of $p_m(\HDG)$ and $X$ must be redistributed over the linear algebra process grid from its  distribution over the DG elements. We have implemented routines for seamlessly inter-converting between a block of vectors distributed over the DG elements to one distributed over the linear algebra process grid, at relatively low communication cost\citep{hu2015dgdft}.  In our experience, this step takes no more than $0.1 \%$ of the total time spent in the CheFSI routine, even for the largest systems considered here. 

The original Chebyshev filtering method presented in \citep{Serial_Chebyshev, Parallel_Chebyshev} employs a QR factorization or the DGKS algorithm \citep{DGKS} for orthonormalization. Here, we have used the faster (but sometimes less stable) Cholesky factorization method instead. We have found this speeds up the orthonormalization by a factor of 2--3 in most cases with no problematic side effects.

With the orthonormalized and filtered block of vectors $\hat{Y}$, the next step is to compute the projection of $H^{\DG}$ onto the occupied subspace: $\hat{H} = \hat{Y}^T (H^{\DG} \hat{Y})$. This step requires the vector block $ \hat{Y}$ distributed  over the linear algebra process grid to be redistributed over the DG elements, so that the action of $H^{\DG}$ on it can be computed. Once again, this data redistribution step takes no more than $0.1 \%$ of the total time spent in the CheFSI routine, even for the largest problems considered here. 

After diagonalizing the projected matrix $\hat{H}$, the resulting block of eigenvectors $Q$ can be used to compute the final results of one CheFSI cycle as $X_{\textrm{new}} = \hat{Y} Q$. 
Eq.~\eqref{eq:densitymatrix} can now be used to compute the diagonal blocks of the density matrix locally on each element, by using the eigenvector coefficients in $X_{\textrm{new}}$. The correponding Ritz values $\Lambda_i$ can be used for adjusting the polynomial filter bounds as well as computing the Fermi energy and occupation numbers. 

\subsubsection{Alignment of eigenvectors with current basis}
\label{subsubsec:wavefun_alignment}
What distinguishes DGDFT from traditional Kohn-Sham DFT solvers is the change of the basis set in each SCF cycle. This change has implications for the way we prepare the starting vectors for CheFSI on every SCF step. Conventionally, the input to CheFSI at the $i^\text{th}$ SCF cycle is chosen to be the approximate invariant subspace computed at the ${(i - 1)}^\text{th}$ SCF 
step \citep{zhou_2014_chebyshev,Serial_Chebyshev}. However, since the basis set changes from one SCF cycle to the next in DGDFT, the eigenvector coefficients computed in a given SCF cycle relative to the basis of that cycle are not applicable to subsequent SCF cycles with different bases. In practice, the change of basis from one SCF cycle to the next becomes  smaller as self-consistency is approached, however, as we show below, the change is sufficiently large to require explicit accommodation to minimize CheFSI iterations.

\begin{figure}
\subfloat{\includegraphics[width=\linewidth]{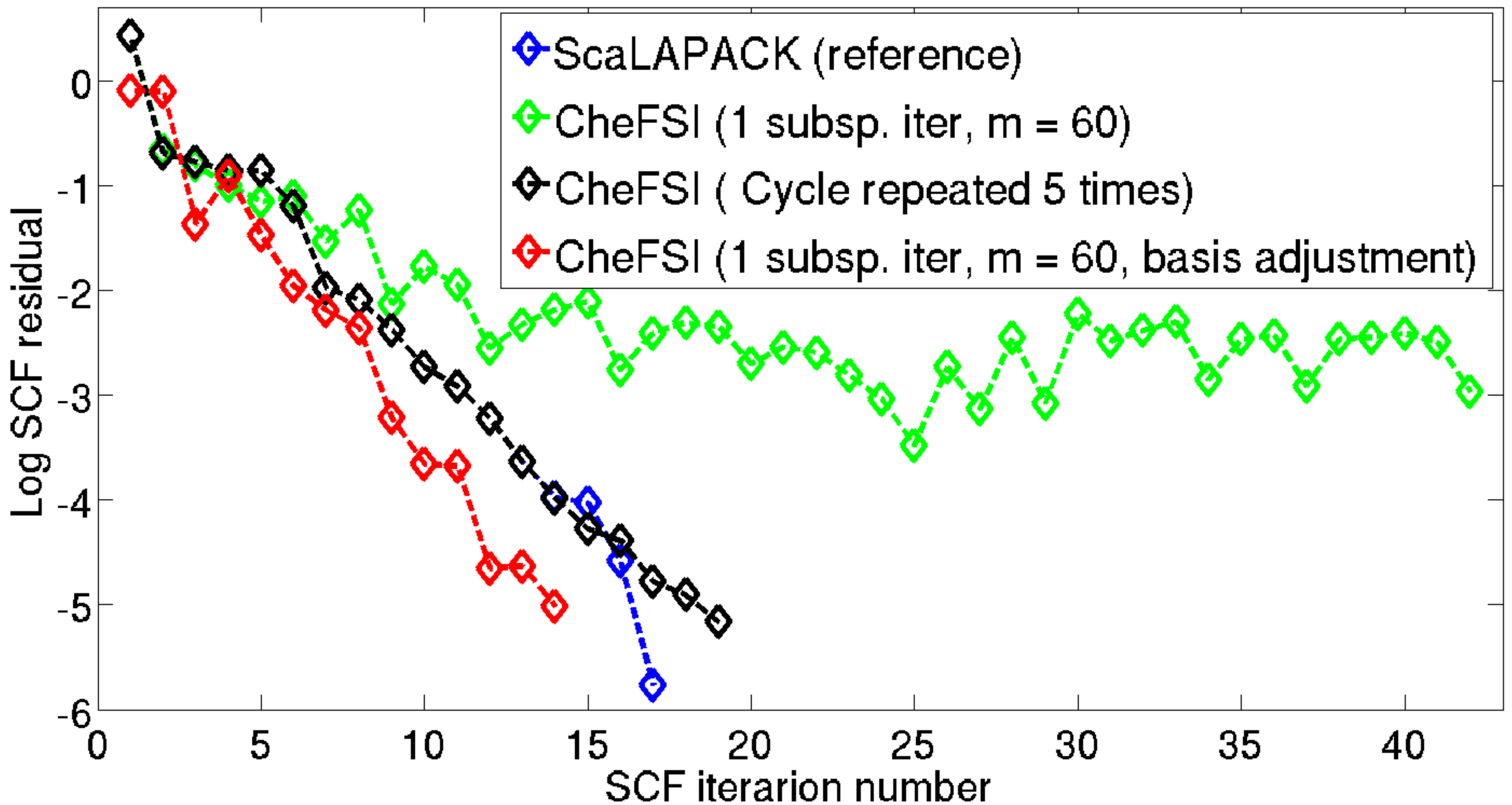}
}
\caption{SCF convergence {of normalized electron density residual} for different variants of CheFSI within DGDFT (naive implementation, multiple cycles, and eigenvector re-alignment to adjust for evolving basis set) for a simple system containing a few hydrogen atoms. Reference ScaLAPACK results are also presented.} 
\label{fig:basis_rotation}
\end{figure}

Figure \ref{fig:basis_rotation} shows that even for a simple system containing a few hydrogen atoms, a naive implementation of CheFSI, which simply uses the eigenvector coefficients computed in the previous SCF cycle as the starting guess for the current SCF cycle, fails to converge in even $45$ iterations (green curve), whereas a direct diagonalization of $H^{\textrm{DG}}$ via ScaLAPACK results in SCF convergence (blue curve) in less than $20$ SCF iterations. To address this problem, we may perform several cycles of CheFSI in every SCF step to compensate for the poor initial approximation provided by the coefficients of the previous SCF step.  This strategy produces results closer to those produced by exact diagonalization in each SCF cycle (black curve.) But repeating the CheFSI cycle multiple times on every SCF step increases the overall computational cost of the method.

Since the basis in a given SCF iteration is distinct from that of the previous, with distinct span, the eigenvectors of the previous SCF iteration cannot be expressed in terms of the basis of the current SCF iteration without approximation. For optimality, we choose the best approximation in the $\ell^2$ 
norm, which by virtue of the orthonormality of the DG basis, is readily obtained by $\ell^2$ projection. Specifically, if $X^{(i)}$ denotes an $N_b \times N_s$ block of vector coefficients (where, $N_b$ denotes the total number basis functions in use and $N_s$ denotes the number of Kohn-Sham states) computed by CheFSI on a given SCF step, and $V^{i}$ denotes an $N_{r} \times N_b$ block of basis vectors corresponding to the ALBs sampled on an $N_r$-dimensional real-space grid (consisting of Gauss-Lobatto integration points for example \cite{lin2012adaptive}), then the starting point for the CheFSI method on SCF step $i+1$ is given by:
\begin{align}
\label{eq:basis_rotation_1}
X^{(i+1)} = (V^{i+1})^T\,V^{i}\,X^{(i)}\,.
\end{align}
Since the ALBs and the eigenvector coefficients $X^{(i)}$ and $X^{(i+1)}$ are all distributed DG-element wise (i.e., $X^{(i)}$ for example, is represented as $M$ blocks $X^{(i)}_K; K = 1,\ldots,M$, stacked column-wise), this becomes:
\begin{align}
\label{eq:basis_rotation_2}
X^{(i+1)} = \sum_{K = 1}^{M} \big((V^{i+1})^T\,V^{i}\big)_K\,(X^{(i)})_K\,.
\end{align}
Further noting that ALBs from different elements are orthogonal to each other due to disjoint supports, we may rewrite the above as:
\begin{align}
\label{eq:basis_rotation_3}
X^{(i+1)}_{K} = (V^{i+1}_K)^T\,V^{i}_K\,(X^{(i)}_K)\,,
\end{align}
with $V^{i}_K$ and $V^{i+1}_K$ denoting the matrix representation of ALBs originating from the element $K$ on SCF steps $i$ and $i+1$, respectively. Eq.~\eqref{eq:basis_rotation_3} can be evaluated locally on each element by means of two matrix-matrix multiplications. As shown by the red curve in Figure \ref{fig:basis_rotation}, this extra step of re-aligning results in SCF convergence with a rate comparable to that of exact diagonalization.

In all the calculations presented here, this extra step of aligning the wavefunction coefficients was always carried out from SCF step 2 onwards. The overhead due to this step is minimal (typically less than $0.1\%$ of the total time spent on a CheFSI cycle) and does not grow with system size since larger systems employ more elements and the re-alignment calculation is carried out locally on each element.

\begin{figure}
\subfloat{\includegraphics[width=0.72\linewidth]{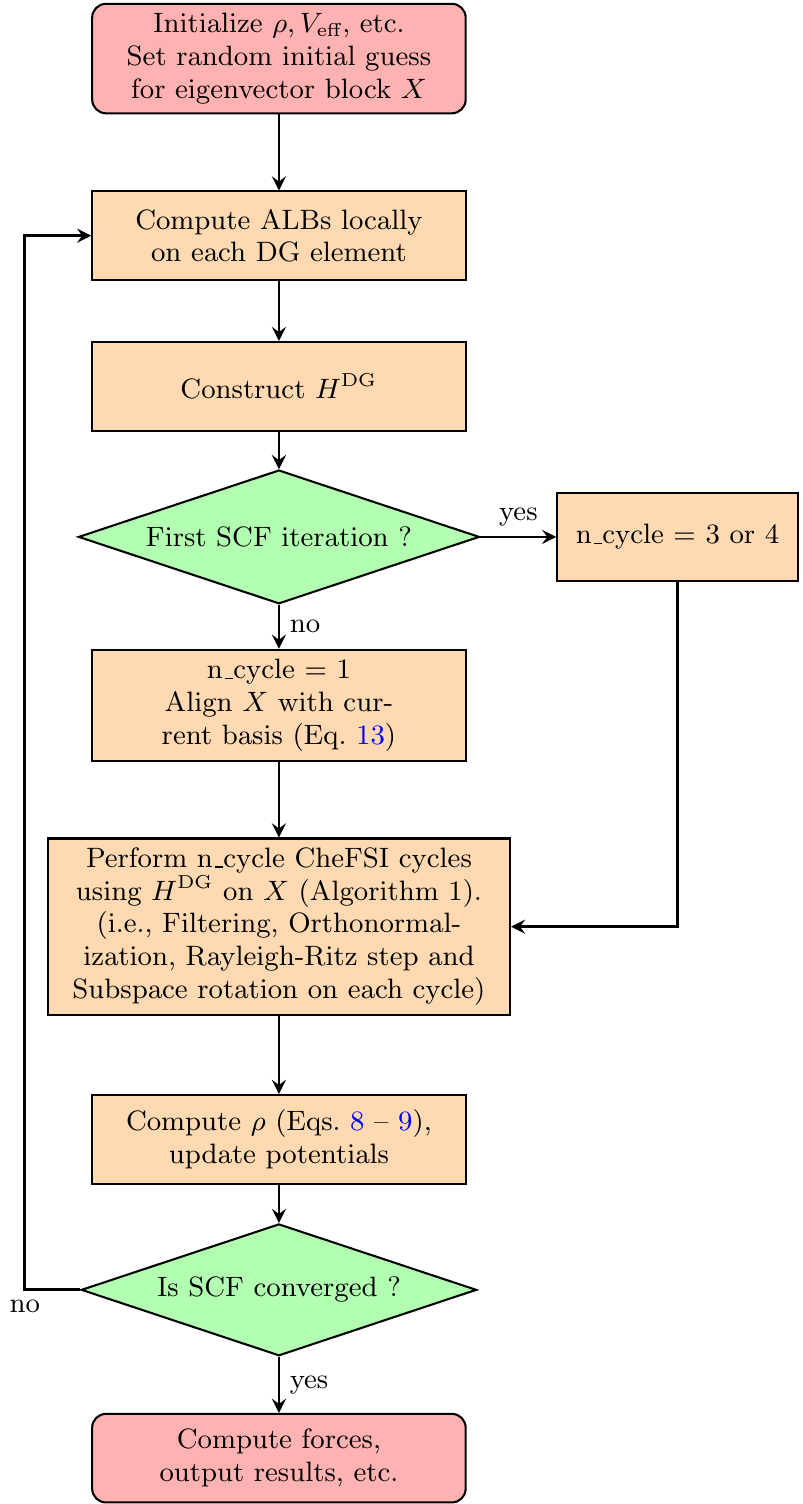}}
\caption{Flowchart depicting the various steps of CheFSI within DGDFT. Typically, $3$ or $4$ CheFSI cycles are applied on the first SCF step when starting from a random guess for the eigenvector block $X$.} \label{fig:flowchart}
\end{figure}

A flowchart summarizing the various steps involved in the DG-CheFSI method is presented in Figure \ref{fig:flowchart}.

\subsubsection{Complexity analysis}
For the purpose of this discussion, we let $N_b$ denote the total number basis functions in use (i.e., $N_b = \sum_{K=1}^{M}J_K$) and we let $N_s$ denotes the number of Kohn-Sham states. Further, we let $N_g$ represent the total number of real-space grid points in use, with $N_g / M$ grid points used for storing each ALB locally within its associated element, with $M$ elements. The quantities $\frac{N_b}{M}$ and $\frac{N_g}{M}$ then correspond to the number of ALBs per element and number of real-space grid points per element, respectively, and are constants for a particular simulation and accuracy level.

As explained above, the CheFSI approach mainly involves the application of the Chebyshev polynomial filter on a block of vectors and subsequent solution of the subspace problem.  Within the DG framework, there is an additional step of aligning the DG coefficients of the Kohn-Sham states from one SCF step to the next. Regardless of the basis set in use (e.g., finite elements, planewaves, or ALBs), the subspace problem solution scales as $\Or(N_b N_s^2 + N_s^3)$ due to the requirement of dense matrix multiplications \citep{Serial_Chebyshev}. 

Let us now focus on the polynomial filtering step. This involves computing the product of the Hamiltonian matrix with the block of Kohn-Sham states. {After the generation process of the ALBs (a step which incurs a memory cost of $\Or({N_g N_b/M})$ on every element),}
%\LL{Why not $N_g N_b/M$? In particular it should not depend on $N_s$} 
the memory cost associated with storage of the coefficients of Kohn-Sham states in terms of the ALBs is 
$\Or\left(\displaystyle{N_bN_s + N_b \frac{N_g}{M}}\right)$. 
This contrasts with the storage cost of $\Or({N_gN_s})$ that would be required by finite differences, finite elements, or planewave methods using the same number of real-space grid points. In this sense, the use of ALBs can be seen as a systematically improvable compressed format for storing the Kohn-Sham states. In practice, the number of ALBs per element $N_b/M$ is at most a few hundreds and this number is usually far exceeded by the number of Kohn-Sham states $N_s$, in large calculations. Further $N_b/N_s$ is typically $2 \sim 20$. Hence, {once the ALBs have been generated}, there is overall less memory cost involved in storing the Kohn-Sham states using the ALBs.

As explained earlier (section \ref{subsubsec:MATVEC}, Eq.~\ref{eq:matvec_2}), multiplying the block sparse matrix $H^{\DG}$ with the block of Kohn-Sham states involves a few dense matrix multiplications of small blocks (of size $\displaystyle \frac{N_b}{M}\times \frac{N_b}{M}$) coming from $H^{\DG}$, with blocks (of size $\displaystyle \frac{N_b}{M}\times N_s$) coming from the coefficients of the Kohn-Sham states, for each DG element. If there are $c_{\mathcal{N}}$ such multiplications to be carried out for each element (this being related to the number of nearest neighbors of elements), the total cost for the application of one step of the polynomial filter is proportional to
\begin{align}
c_{\mathcal{N}} \bigg(\frac{N_b}{M}\bigg)^2 N_s M = \Or(N_s M)\,,
\label{eq:filt_scaling_DG}
\end{align}
since $\frac{N_b}{M}$ is a constant.
Splitting this calculation into the respective $M$ elements therefore incurs a computational cost of $\Or(N_s )$ on every element. Note that, in contrast, this computation using finite differences or finite elements would have a complexity of $\Or(N_sN_g)$ in total and is likely to incur a greater computational cost.

Finally, the step of aligning the Kohn-Sham states with the current basis set on every SCF step (Eq.~\ref{eq:basis_rotation_2}) incurs a cost that is proportional to
\begin{align}
\bigg(\frac{N_b}{M}\bigg)^2 \frac{N_g}{M} + \bigg(\frac{N_b}{M}\bigg)^2 N_s = \Or(N_s)\,,
\end{align}
on every element. 

In contrast to the computational complexity of the various steps involved in the CheFSI approach, direct diagonalization of $\HDG$ involves a computational complexity of $\Or(N_b^3)$ while the PEXSI approach involves a cost of $\Or\bigg(\big(\frac{N_b}{M}\big)^3M^{\alpha_D}\bigg)$ (with $\alpha_D = 1.0, 1.5, 2.0$ for one-, two-, and three-dimensional systems, respectively). Direct diagonalization is more computationally intensive while the PEXSI approach results in a larger prefactor, because of which both  methods result in longer wall times to solution compared to CheFSI for the full range of system sizes considered here.

\section{Results}
\label{sec:results}
\begin{figure}
\subfloat[Bulk $\textrm{Li3D}$ system containing $318$ atoms. \label{subfig:Li3D}]{
  \includegraphics[width=0.55\linewidth]{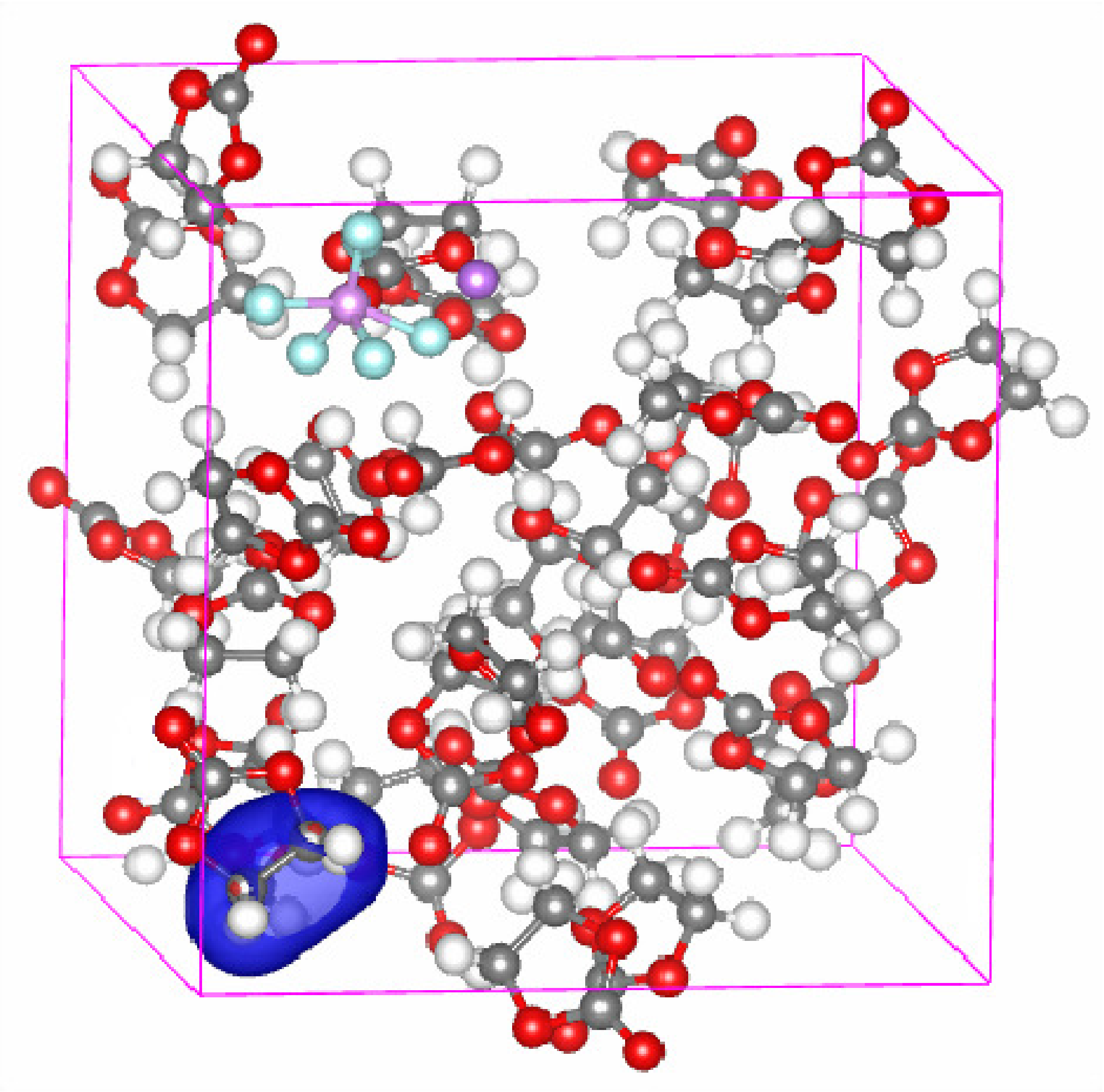}
}\\
\subfloat[$\textrm{Graphene2D}$ system containing $180$ atoms. \label{subfig:Graphene2D}]{
  \includegraphics[width=0.55\linewidth]{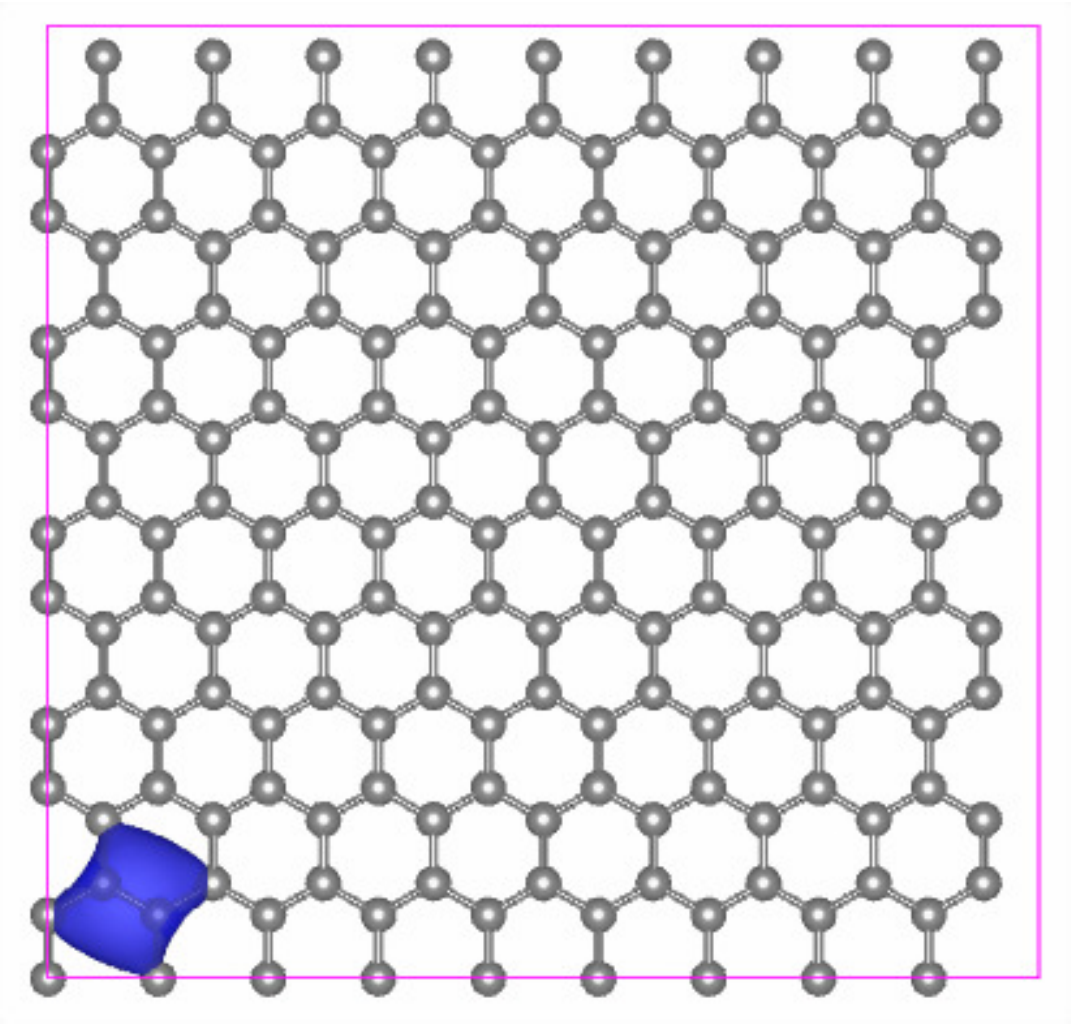}
}
\caption {Prototype 2D and 3D systems used for the computations in this work. Larger sized systems were obtained by periodic replication of these unit cells. Iso-surface of the first ALB from one of the DG elements of these systems is also shown.}
\label{fig:prototype_systems}
\end{figure}

In this section, we investigate the parallel scalability of the CheFSI method and compare its performance with the existing ScaLAPACK and PEXSI methods in DGDFT. Two prototypical systems have been used for our calculations. The first, referred to as \emph{Li3D}, consists of a three-dimensional bulk lithium-ion electrolyte system originating from the design of energy storage devices.  Atoms of hydrogen, lithium, carbon, phosphorus, oxygen, and fluorine, numbering $318$ in total, are present in this system. The second, referred to as \emph{Graphene2D}, consists of a two-dimensional sheet of graphene containing 180 carbon atoms. These systems were chosen for their technological relevance as well as the fact that KS-DFT calculations on large samples of these systems can be challenging. Figure \ref{fig:prototype_systems} shows the \emph{Li3D} and \emph{Graphene2D} systems along with the first ALB from one of the DG elements of these systems.

In order to be able to work with larger system sizes, we have employed multiple unit cells of these systems replicated along the coordinate axes. Thus, $\textrm{Li3D}_{1 \times 2 \times 2}$ for example, refers to a system in which the $318$-atom unit cell has been replicated along Y and Z directions to produce a $1272$-atom bulk system; and similarly, $\textrm{Graphene2D}_{4 \times 4}$ refers to a graphene sheet containing $2880$ atoms.

\begin{figure}
\subfloat{\includegraphics[width=\linewidth]{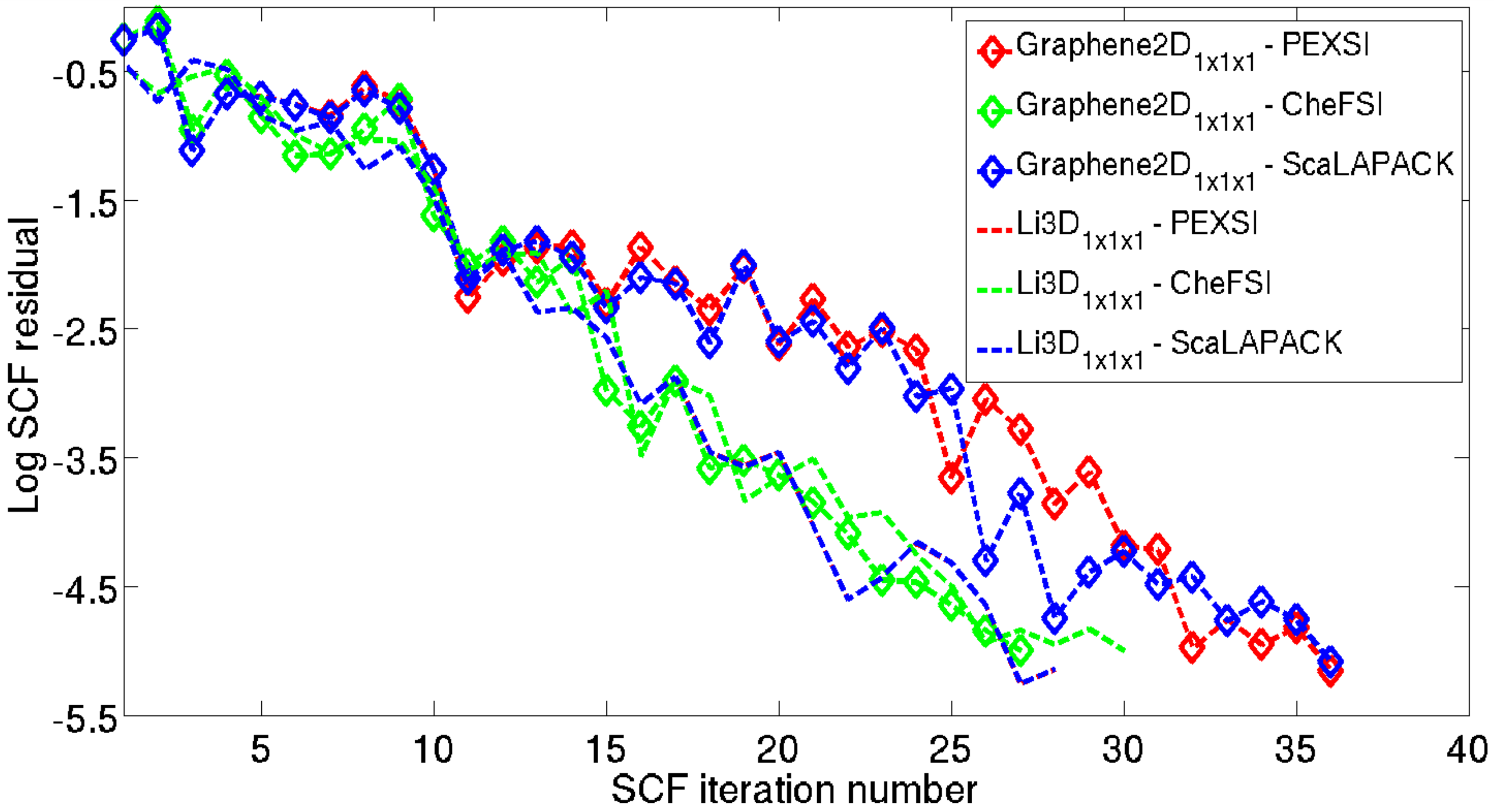}
}
\caption{SCF convergence {of normalized electron density residual} for two prototypical systems using ScaLAPACK diagonalization, PEXSI, and CheFSI methods.}
\label{fig:scf_conv}
\end{figure}

In what follows, we shall consider the \emph{time to solution}. For the CheFSI and ScaLAPACK diagonalization methods, this will refer to the wall clock time that these methods require to compute the eigenvalues and eigenvectors of $H^{\DG}$ as well as the diagonal blocks of the density matrix (via Eq.~\eqref{eq:densitymatrix}), during a general SCF cycle. For the PEXSI method, it will refer to the wall clock time that is required to compute directly the density matrix corresponding to the sparsity pattern of $H^{\DG}$. In order to have a fair comparison between the methods, it is important to ensure that the three methods show the same convergence rate over multiple SCF cycles. This then allows the comparison between the methods to be carried out with reference to the time to solution for one SCF cycle. Accordingly, we have adjusted the Chebyshev polynomial filter order as well as the various parameters used in PEXSI (such as the number of poles and the number of chemical potential iterations), so that these methods converge at least as fast as the reference ScaLAPACK calculations. Figure \ref{fig:scf_conv} shows the convergence of all the three methods for the prototype systems in use here.

For most of the calculations described here, the polynomial filter order used was between 80 and 100.  However, these employed relatively hard pseudopotentials \citep{GTH_relativistic}; lower filter orders may be expected to suffice for softer pseudopotentials. Additionally, in practical molecular dynamics and geometry optimization calculations, fewer SCF cycles are likely to be required for the CheFSI method on every electronic relaxation step, since the method will be able to make use of wavefunction extrapolation {(with re-alignment to account for the  evolving basis set)}.  Thus, the performance of CheFSI in the context of MD simulations may be expected to improve still further relative to the results of static calculations, as presented here.

We have used the local density approximation for the exchange-correlation functional with the parametrization described in \citep{GTH_pseudoptential}. Hartwigsen-Goedecker-Teter-Hutter\\ pseudopotentials \citep{GTH_pseudoptential, GTH_relativistic} are employed to remove inert core electrons from the computations. We have typically employed $100-120$ additional states in most calculations to accommodate partial occupation. SCF convergence was accelerated by means of Pulay's scheme \citep{pulay_mixing} or its periodic variant \citep{periodic_pulay}, and an electronic temperature of $300$ K was used in Fermi-Dirac occupation. To attain chemical accuracy {(i.e., error in  the total energy less than $10^{-3}$ Ha/atom relative to the fully converged result; additionally we also ensured that the error in the atomic forces are less than $10^{-3}$ Ha/Bohr relative to the fully converged result)}, 
%the scenario in which the errors in total energy per atom and the atomic forces are less than $10^{-3}$ when compared to well converged plane-wave code results -- in practice this is sufficient for reliable ab initio molecular dynamics simulations)}, 
%\LL{I wonder whether our energy per atom accuracy is $10^{-4}$. Also I guess we might need to avoid using ``chemical accuracy'' since this is an accuracy measured against experiment rather than the numerical accuracy which can be arbitrarily accurate.} 
the 318-atom bulk Li3D system was partitioned into $4\times4\times4$ elements, with 200 ALBs per element, giving $\sim$40 ALBs per atom. Likewise, the 180-atom Graphene2D system was partitioned into $1\times6\times6$ elements, with 120 ALBs per element, giving 24 ALBs per atom.

All calculations described here were performed on the Edison platform at
the National Energy Research Scientific Computing (NERSC) center.
Edison has 5462 Cray XC30 nodes. Each node has 64 GB of memory and 24 cores partitioned
among two Intel Ivy Bridge processors, running at 2.4GHz. Edison employs a Cray Aries high-speed interconnect with Dragonfly topology for inter-node communication.

\subsection{Scaling performance}
\label{subsec:scaling_performance}
We first investigate the strong scaling performance of CheFSI within the DG framework and compare it against that of PEXSI and ScaLAPACK. For this, we consider the systems $\textrm{Li3D}_{2 \times 2 \times 2}$ ($2544$ atoms, $4536$ Kohn-Sham states) and $\textrm{Graphene2D}_{6 \times 6}$ ($6480$ atoms, $13080$  Kohn-Sham states).

\begin{figure}
\subfloat[$\textrm{Li3D}_{2 \times 2 \times 2}$ system (2544 atoms). \label{subfig:Li_strong}]{
  \includegraphics[width=\linewidth]{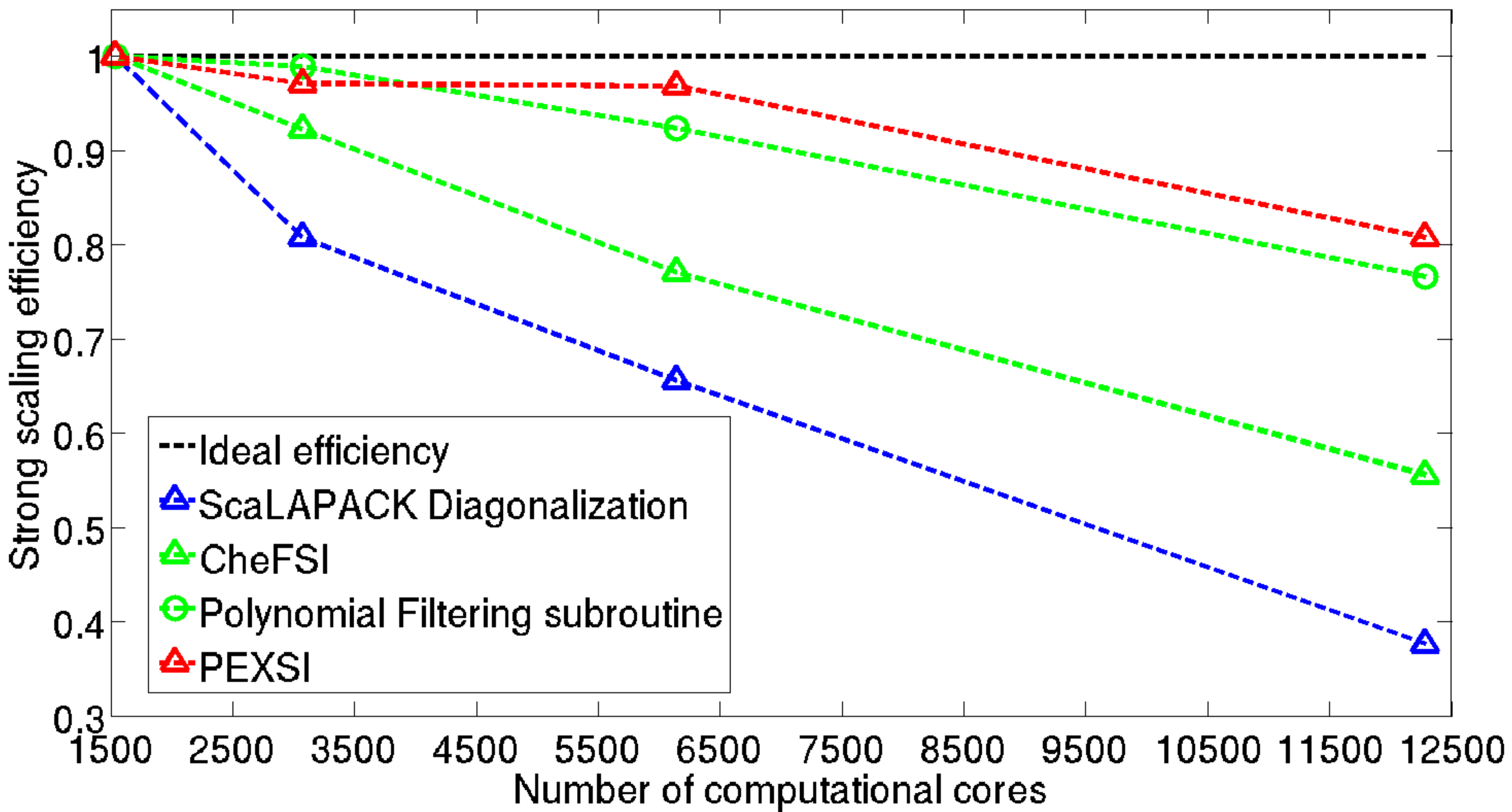}
}\\
\subfloat[$\textrm{Graphene2D}_{6 \times 6}$ system (6480 atoms). \label{subfig:Graphene_Strong}]{
  \includegraphics[width=\linewidth]{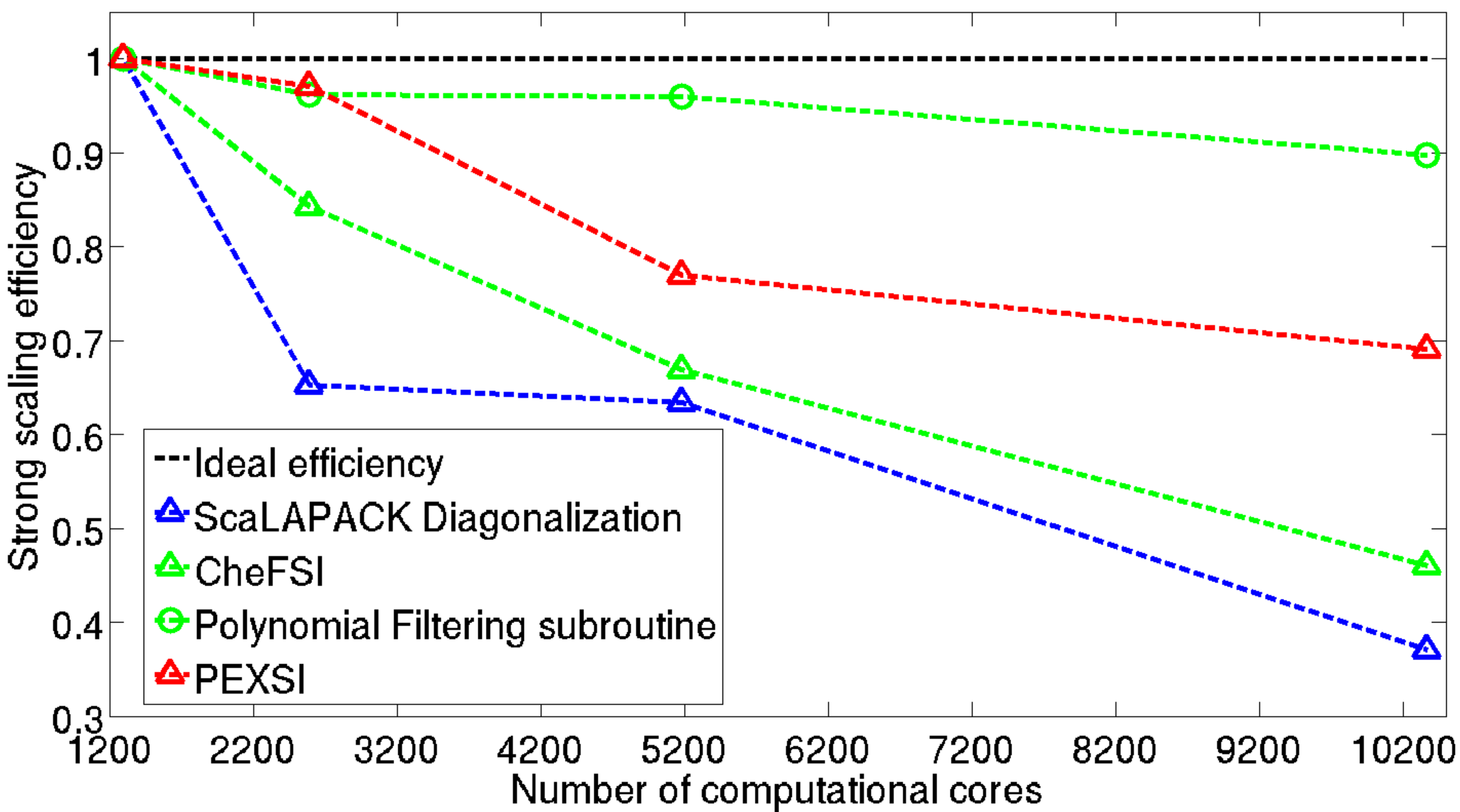}
}
\caption{Strong scaling efficiency of CheFSI in DGDFT, compared against PEXSI and direct ScaLAPACK diagonalization. Scaling performance of the filtering routine is also shown.}
\label{fig:strong_scaling}
\end{figure}

Figure \ref{fig:strong_scaling} shows the wall time to solution (per SCF iteration) vs.\ number of computational cores employed. 
From the figures, it is evident that the overall strong scaling performance of CheFSI lies in between that of PEXSI and direct ScaLAPACK diagonalization. For the $\textrm{Li3D}_{2 \times 2 \times 2}$ system, the performance using $12,288$ cores is at about $56$ \% efficiency (measured against the result from $1500$ cores); while for the $\textrm{Graphene2D}_{6 \times 6}$ system, using $10,368$ processors, it is at about $46$ \% efficiency (measured against the result from $1200$ cores). It is interesting to note, however, that the strong scaling performance of the filtering routine by itself is nearly ideal, remaining close to $80$ \% efficiency for the  $\textrm{Li3D}_{2 \times 2 \times 2}$ case and at about $90$ \%  efficiency for the $\textrm{Graphene2D}_{6 \times 6}$ case. In particular, the performance of the filtering routine is better in the 2D system due to fewer neighboring elements and correspondingly less communication required. The overall scaling performance of CheFSI, therefore, is limited 
by the performance of the subspace problem solution, whenever the total time for this step forms a significant fraction of the total CheFSI time. For the systems here, the subspace problem solution time was about $33$ \% of the total CheFSI time for the $\textrm{Li3D}_{2 \times 2 \times 2}$ system using $12,288$ cores, and $57$ \% of the total CheFSI time for the $\textrm{Graphene2D}_{6 \times 6}$ system  using $10,368$ cores. The larger fraction of time spent on the subspace problem helps explain why the overall scaling performance of CheFSI is somewhat lower for the 2D case here. These observations  suggest possible avenues for further improvement of the overall scaling performance of CheFSI.

\begin{figure}
\subfloat{\includegraphics[width=\linewidth]{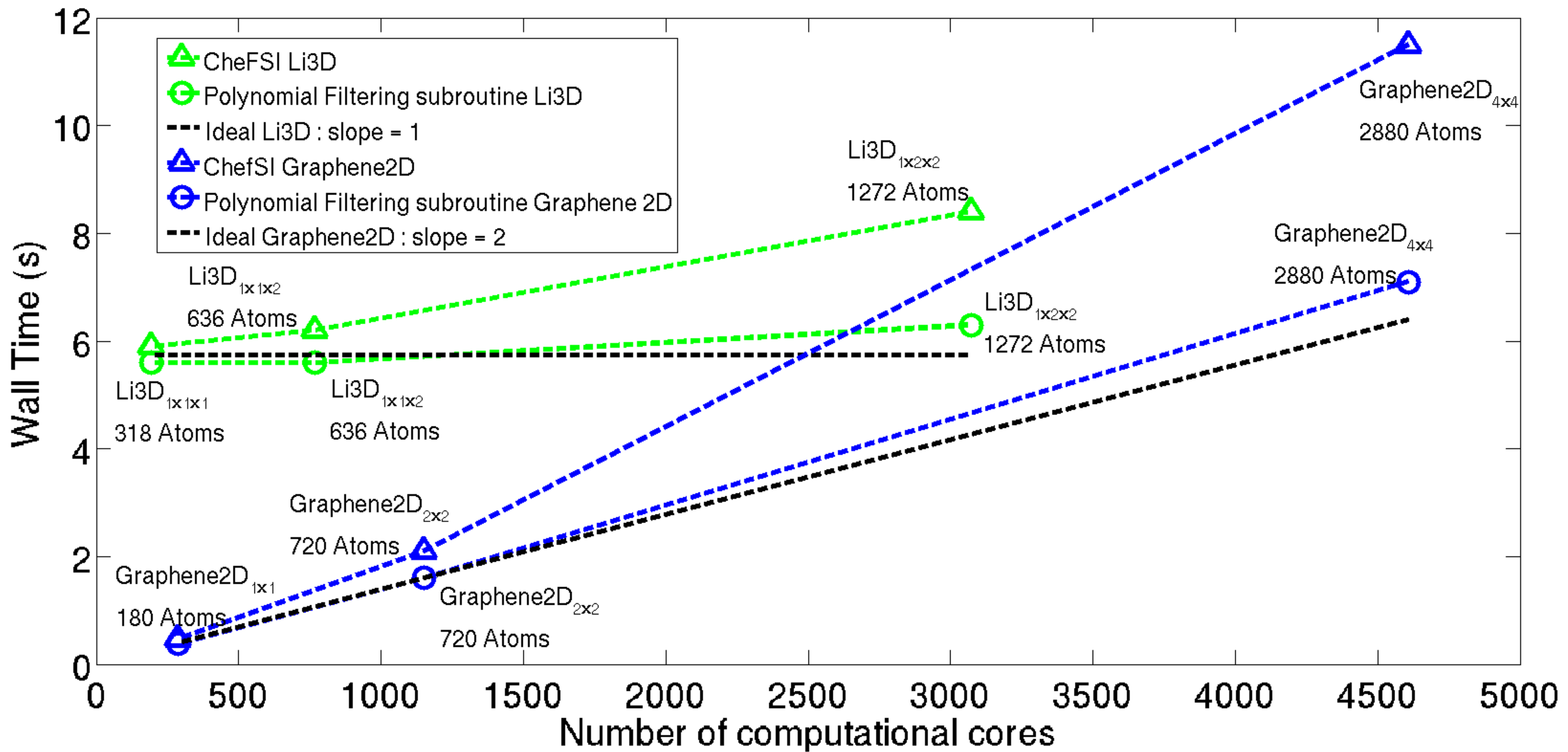}
}
\caption{Weak scaling performance of CheFSI in DGDFT. Performance of the filtering routine is also shown.}
\label{fig:weak_scaling}
\end{figure}

Next, we investigate the weak scaling performance of CheFSI within DGDFT, i.e., the performance with increasing system size. We investigate the following systems in 3D: $\textrm{Li3D}_{1 \times 1 \times 1}, \textrm{Li3D}_{1 \times 1 \times 2}$, and $\textrm{Li3D}_{1 \times 2 \times 2}$. The system sizes have been doubled successively and as a result, the number of Kohn-Sham states involved (approximately) is doubled as well. In 2D, we investigated the systems: $\textrm{Graphene2D}_{1 \times 1}, \textrm{Graphene2D}_{2 \times 2}$, and $\textrm{Graphene2D}_{4 \times 4}$. For these cases, the system sizes have been quadrupled successively and as a result, the number of Kohn-Sham states involved (approximately) is quadrupled as well. As a measure of weak scaling performance, the wall clock time to solution (per SCF iteration) is shown in Figure \ref{fig:weak_scaling}. For each system, the number of computational cores was quadrupled successively as sizes were doubled ($\textrm{Li3D}$) and quadrupled ($\textrm{Graphene2D}$) successively.

On increasing the system size $n$ fold, the number of DG elements used for the calculation has to increase by the same factor to keep each local calculation manageable and to maintain the same level of accuracy of solution. Since the number of Kohn-Sham states involved also increases $n$ fold, there is an overall $n^2$ factor increase in the time required for applying the Chebyshev polynomial filter to all the states involved (Eq.~\ref{eq:filt_scaling_DG}). Hence, if the number of computational cores used for doing the larger calculation is increased by a factor of $s$, the expected wall time for applying the Chebyshev polynomial filter will change by a factor of $n^2/s$. This observation should also hold for the total CheFSI time, as long as the time for solution of the subspace problem forms a small fraction of the total CheFSI time. Thus, for the $\textrm{Li3D}$ systems, the wall-times for each system should ideally remain constant ($n =2, s = 4$) while for the $\textrm{Graphene2D}$ systems, an increase by a factor of $4$ should be observed ($n =4, s = 4$). Figure \ref{fig:weak_scaling} shows that this expectation holds reasonably well for the overall CheFSI time, and particularly well for the Chebyshev polynomial filter application time. For the $\textrm{Li3D}$ systems, the weak scaling efficiency of CheFSI is about $70$ \% using  $3072$ cores while it is about $65$ \% using $4608$ cores for the $\textrm{Graphene2D}$ systems. The performance of the polynomial filter application routine for both these systems is close to $90$ \%. 
These results demonstrate again the critical importance of an efficient, well scaling subspace solution as system size increases beyond a few thousand atoms.
\subsection{Benchmark calculations}
\label{subsec:benchmark_calculations}
As the final test of computational efficiency, we study the performance of CheFSI on large benchmark systems and compare the wall time to solution between CheFSI, direct ScaLAPACK diagonalization, and PEXSI. We choose the $\textrm{Li3D}_{3 \times 3 \times 3}$ and   $\textrm{Graphene2D}_{8 \times 8}$ systems for this study. $13,824$ computational cores were used for both systems. The results are shown in Table \ref{tab:benchmark_walltime}.
\begin{table}[h]
\centering
\begin{tabular}{  c  c  c  c  }
\hline\hline			
System & ScaLAPACK & PEXSI & CheFSI \ \\
\hline  
$\textrm{Li3D}_{3 \times 3 \times 3}$ &  &  & \\
 $8,586$ atoms & $3323$ & $3784$ & $170$ \\
 $\sim 15,000$ states  &  &  & \\\hline
$\textrm{Graphene2D}_{8 \times 8}$ &  &  & \\
 $11,520$ atoms & $2473$ & $426$ & $105$ \\
 $\sim 23,200$ states  &  &  & \\\hline
\end{tabular}
\caption{Solution wall times per SCF step (rounded to nearest second) for direct ScaLAPACK diagonalization, PEXSI, and CheFSI on $13,824$ computational cores for two large systems.}\label{tab:benchmark_walltime}
\end{table}

The results show that CheFSI is by far the fastest of all the three approaches (up to more than an order of magnitude faster), particularly for the bulk system.  Even for the two-dimensional material system, a geometry in which PEXSI is known to perform particularly well, CheFSI is able to outperform with the same number of cores. Due to the good scalability properties of PEXSI, the wall time for the $\textrm{Graphene2D}_{8 \times 8}$ system can be brought down to be comparable to CheFSI (using $55,296$ cores for instance), but overall CheFSI remains more economical in terms of computational resources used (total CPU-hours, for example). We have also observed that the timing results remain favorable for CheFSI for smaller systems, such as those used in the scaling performance studies.

In order to obtain an estimate of the SCF wall times achievable with the DGDFT-CheFSI framework on large-scale computational platforms, we studied the $\textrm{Li3D}_{3 \times 3 \times 3}$ and $\textrm{Graphene2D}_{8 \times 8}$ systems using 55,296 computational cores. The results are shown in Table \ref{tab:scf_walltime}. 
\begin{table}[h]
\centering
\resizebox{0.75\columnwidth}{!}{
\begin{tabular}{ c  c  c  c  c  }
\hline\hline			
System & ALB  & Hamiltonian & CheFSI & Total SCF \\
  & Generation & update & (filtering) & time \ \\ \hline
$\textrm{Li3D}_{3 \times 3 \times 3}$  & $11$ & $3$ & $76\,(36)$ & $90$\ \\
$\textrm{Graphene2D}_{8 \times 8}$ & $5$ & $4$ & $66\,(16)$ & $75$\\
\hline
\end{tabular}
}
\caption{Wall times for various stages of the SCF cycle (rounded to nearest second) with the DGDFT--CheFSI approach  for two large systems using $55,296$ computational cores. The numbers in parentheses indicate the wall time spent on the filtering step.}\label{tab:scf_walltime}
\end{table}

It is apparent from the results in Table \ref{tab:scf_walltime} that for these systems, the largest fraction of the total SCF time is spent on the solution of the subspace problem. Thus the cubic computational complexity associated with the solution of the subspace problem starts to dominate as the system size grows larger, beyond a few thousand atoms in the present case. 

\section{Conclusion}
\label{sec:conclusion}
We have used Chebyshev polynomial filtered subspace iteration (CheFSI) within the Discontinuous Galerkin method to enable large-scale first principles simulations of a wide variety of materials systems using Density Functional Theory. Due to a number of attractive features of the DG Hamiltonian matrix, the implementation of CheFSI within the Discontinuous Galerkin framework allows the computation of the Kohn-Sham eigenstates of the Hamiltonian to be carried out in a highly efficient and scalable manner. By virtue of the limited spectral width of the DG Hamiltonian matrix, relatively low polynomial orders suffice, reducing the number of matrix-vector multiplies required; while the block-sparse structure of the DG Hamiltonian facilitates efficient, parallel implementation of each multiply. In addition, the strict locality and orthonormality of the adaptive local basis facilitates realignment of eigenvector coefficients from one SCF step to the next, as the basis is optimized on-the-fly at each step. Taken together, these advantages yield an accurate, systematically improvable electronic structure method, applicable to metals and insulators alike, capable of simulating thousands of atoms in tens of seconds per SCF iteration on large-scale parallel computers.

In the near future, we aim to carry out large-scale quantum molecular dynamics simulations of various materials systems using the DGDFT-CheFSI technique. Of particular interest to us are accurate simulations of the solid-electrolyte interphase (SEI) layer in lithium-ion batteries, and we anticipate that the new  methodology will enable accurate simulations of unprecedented size.

While the current methodology can simulate a few thousand atoms in a few tens of seconds per SCF iteration with planewave accuracy, to reach further still, to 10,000 atoms or more with comparable efficiency, will require a substantially more efficient and scalable solution of the subspace problem. We aim to address this issue in future work. One possible avenue for making this step more scalable is to replace the use of Cholesky factorization and eigensolution (for the Rayleigh-Ritz step) with operations which involve only parallel dense matrix multiplications in the occupied subspace. Parallel dense matrix multiplication tends to scale more favorably and therefore stands to relieve the scalability bottleneck in the current approach. {Yet another, more radical possibility would be to dispense with the CheFSI methodology completely, thus avoiding the Rayleigh-Ritz step. Computational techniques such as FEAST \citep{polizzi2009density} or spectrum slicing \citep{schofield2012spectrum} might be used to compute the spectrum of $\HDG$ instead. However, compared to more conventional methods like CheFSI, these techniques are likely more suitable for the next generation of computing platforms \citep{levTor15}. An interesting avenue for future work, therefore, would be to investigate whether such techniques can be made to yield significant performance benefits on current parallel computing platforms for physical systems of the types and sizes considered here.}

{Finally, comparison of the performance of DGDFT-CheFSI with other massively parallel electronic structure codes, such as Qbox \citep{gygi2005large, gygi2008architecture} for example, is another interesting avenue for research, which the authors are pursuing presently.}  

\section*{Acknowledgments}\label{acknowledgments}
This work was performed, in part, under the auspices of the U.S.~Department of Energy by Lawrence Livermore National Laboratory under Contract DE-AC52-07NA27344. Support for this work was provided through Scientific Discovery through Advanced Computing (SciDAC) program funded by U.S.~Department of Energy, Office of Science, Advanced Scientific Computing Research and Basic Energy Sciences (A.S.B., L.L., W.H., C.Y., and J.E.P), and by the Center for Applied Mathematics for Energy Research Applications
(CAMERA), which is a partnership between Basic Energy Sciences and Advanced Scientific Computing Research at the U.S Department of Energy (L. L. and C. Y.). The authors thank the National Energy Research Scientific Computing (NERSC) center for making computational resources available to them. A.S.B. would like to thank Meiyue Shao (Lawrence Berkeley Lab) for informative discussions and for his help with improving the presentation of the manuscript. The authors would also like to thank the anonymous reviewers for their comments which helped in improving the manuscript.
\bibliography{DG_Cheby}

\end{document}